\RequirePackage{xcolor}
\RequirePackage{ifpdf}
\documentclass[12pt,letterpaper]{JHEP3}

\usepackage{amscd,amsmath,amssymb,amsfonts,xspace,mathrsfs}
\usepackage[utf8]{inputenc}

 \usepackage{tikz-cd}
 \usepackage{amssymb}
 \usetikzlibrary{cd}
 \usetikzlibrary{arrows,intersections}
\usepackage{xargs,bm}                      
\usepackage{nccmath}
\usepackage{bm}
\usepackage{caption}
\newcommand{\be}{\begin{equation}}
\newcommand{\ee}{\end{equation}} 
\newcommand{\bes}{\begin{equation*}}
\newcommand{\ees}{\end{equation*}}
\newcommand{\sgn}{\mathrm{sgn}}

 \usepackage{enumerate}

\newcommand{\CA}{\mathcal{A}} 
\newcommand{\CB}{\mathcal{B}} 
  
\newcommand{\cD}{\mathcal{D}}
\newcommand{\CE}{\mathcal{E}}  
\newcommand{\CF}{\mathcal{F}} 
 
\newcommand{\CH}{\mathcal{H}}

\newcommand{\CK}{\mathcal{K}}
\newcommand{\CL}{\mathcal{L}} 
\newcommand{\CM}{\mathcal{M}} 
\newcommand{\CN}{\mathcal{N}}
\newcommand{\CO}{\mathcal{O}} 

\newcommand{\CQ}{\mathcal{Q}}

\newcommand{\CS}{\mathcal{S}}

\newcommand{\BC}{\mathbb{C}}

\newcommand{\BF}{\mathbb{F}}
\newcommand{\BR}{\mathbb{R}}
\newcommand{\BH}{\mathbb{H}}
\newcommand{\BM}{\mathbb{M}}
\newcommand{\BL}{\mathbb{L}}

\newcommand{\BP}{\mathbb{P}}
\newcommand{\BZ}{\mathbb{Z}}
\newcommand{\BQ}{\mathbb{Q}}

\newcommand{\SFO}{\mathsf{O}}

\newcommand{\bra}{\langle}
\newcommand{\ket}{\rangle}

\newcommand{\eps}{\epsilon}

\newcommand{\bfb}{{\boldsymbol b}}

\newcommand{\bfl}{{\boldsymbol l}}

\newcommand{\bfv}{{\boldsymbol v}} 
\newcommand{\bfx}{{\boldsymbol x}}
\newcommand{\bfy}{{\boldsymbol y}}
\newcommand{\bfrho}{{\boldsymbol \rho}}

\newcommand{\bfk}{{\boldsymbol k}}
\newcommand{\bfz}{{\boldsymbol z}}
\newcommand{\bfmu}{{\boldsymbol \mu}}

\newcommand{\bfS}{{\boldsymbol S}}

\newcommand{\TT}{\tau}
\newcommand{\SU}{\mathrm{SU(2)}}

\newcommand{\SOT}{\mathrm{SO(3)}}

\title{Donaldson-Witten theory, surface operators and mock modular forms 
}
\author{Georgios Korpas \\
{\it School of Mathematics, Trinity College, Dublin 2, Ireland}\\
{\it Hamilton Mathematical Institute, Trinity College, Dublin 2, Ireland}\\

\vspace*{2mm} {\tt E-mail:
\email{george.korpas@maths.tcd.ie}
}
}

\abstract{ We revisit the $u$-plane integral of the topologically twisted $\CN=2$ super Yang-Mills theory, the Donaldson-Witten theory, on a closed four-manifold $X$ with embedded surfaces that support supersymmetric surface operators. This integral mathematically corresponds to the generating function of the ramified Donaldson invariants of $X$. By including a $\overline{\CQ}$-exact deformation to the $u$-plane integral we are able to re-express its integrand in terms of a total derivative with respect to an indefinite theta function, a special kind of mock modular form. We show that for specific K\"ahler surfaces of Kodaira dimension $-\infty$ the integral localizes at the cusp at infinity of the Coulomb branch of the theory.}

\preprint{TCD-MATH 18--12}

\begin{document}

\section{Introduction} 
Topological quantum field theories (TQFTs) that arise by twisting various supersymmetric gauge theories have played and continue to play a prominent role in the field of differential topology over the past 30 years. A great deal of effort has been put into understanding the topology of (mainly low dimensional) manifolds and the study of these topological field theories has led to a physical interpretation of certain (topological, diffeomorphism, etc.) invariants of manifolds in terms of partition functions and correlation functions of the topological quantum field theories studied on them. Some typical examples of such invariants are the Donaldson invariants \cite{DONALDSON1990257, Witten:1988ze}  and Seiberg-Witten invariants \cite{Witten:1994cg} of four manifolds which are obtained from gauge theoretic constructions or Gromov-Witten invariants \cite{GW} and Donaldson-Thomas invariants \cite{DT} for Calabi-Yau three-folds, to name a few, while more recently a particular TQFT (the GL-twist of the $\CN=4$ SYM theory) has provided a physical interpretation of various aspects of the geometric Langlands program \cite{Kapustin:2006pk}. 

After Witten's fundamental work \cite{Witten:1988ze} the physics community took the subject by storm especially due to the apparent connections to supersymmetric gauge field theories.  Usually it is very hard to obtain exact results for such theories but sometimes it is possible to do so in their so-called BPS sectors, that is sectors whose quantum corrections are well controlled. Various BPS quantities, for instance correlators of BPS operators, can be computed exactly therefore because a semi-classical approximation is enough. These quantities often appear as generating functions of some topological invariants of the space that the theory is evaluated on. A very well known example is that of the partition function of the $\CN=4$ topologically twisted super Yang-Mills theory, the Vafa-Witten theory, on a four-manifold $X$ which computes the Euler numbers of the moduli space of sheaves on $X$ \cite{Vafa:1994tf}.

Specifically, and of main focus in the present paper, we are interested in Donaldson theory with embedded surfaces. But let us recall first the usual Donaldson theory which is defined in terms of gauge theory evaluated on a four-manifold with gauge group $\mathrm{SU}(2)$ or ${\rm SO}(3)$. More concretely, let $P$ be a principal $\mathrm{SU}(2)$-bundle\footnote{We choose $\mathrm{SU}(2)$ for simplicity.} over a four-manifold $X$ with Riemannian metric $g$. For $b_2(X) =\mathrm{dim}(H_2(X,\BR))$ we denote by $b_2^+(X)$ the dimension of its positive definite subspace and by $b_2^{-}(X) $ the dimension of its negative definite subspace\footnote{Equivalently, the dimensions of the positive and negative harmonic two-forms respectively.}. Denote by $E$ the associated bundle to $P$. To this bundle we can assign a lot of connections\footnote{Connections on $E$ are induced from connections on $P$.} but we want to consider the irreducible ones $A$ that solve the anti-self-dual (ASD) equation $*F=-F$,  where $F=d_AA$ is the curvature of $A$. In addition, we denote by $\CM_k$ the (appropriately compactified) moduli space of solutions of the ASD equation, with $k = c_2(E)$. Such solutions physically correspond to instantons.  Donaldson invariants, which are smooth structure invariants of $X$, are obtained by calculating integrals  of various differential forms over the moduli space $\CM_k$. To be more precise, let $H_*(X,\BQ)$ be the rational homology ring of $X$. Then we can construct the universal (instanton) bundle as $\CE := E \times \CM_k \to X \times \CM_k$ with an associated universal connection $\CA$. Then, we would like to ``integrate out" the dependence on the homology classes of the four-manifold X. This goes through with the slant product (Donaldson map) which is defined as the map $\mu : H_i(X,\BQ) \to H^{4-i}(\CM_k,\BQ)$ such that for a class $\beta \in H_i(X,\BQ)$ we get $\mu(\beta) = (c_2(\CE)- c_1(\CE)^2/4 )/\beta$. There are cases where the universal bundle does not exist though and then the construction is carried through using the endomorphism bundle $\mathrm{End}(E)$ instead. Next, let us denote by $s$ the dimension of $\CM_k$ which can be computed with the Atiyah-Singer index theorem. It can be shown that for gauge group  ${\rm SU}(2)$  \cite{Donaldson90}
\be
s = 8c_2(E)-3(1-b_1(X) +b_2^+(X)).
\ee 
Then, for classes $\bfx_1, \ldots, \bfx_m \in H_2(X,\BQ)$ and for a point class $p \in H_0(X,\BQ)$  the Donaldson invariant of degree $s$ is defined as\footnote{If the base four-manifold $X$ has $b_2^+=1$ then the definition of Donaldson invariants depends on the choice of a period point $J \in H^2(X,\BR)$ as well. The reason for this is because for four-manifolds with $b_2^+=1$ we can have reducible solutions to the ASD equation corresponding, in the case of vector bundles over K\"ahler surfaces, to semi-stable bundles. These solutions occur in one parameter families that we call walls (see Section \ref{WCformula}). Then, a period point $J \in H^2(X,\BR)$ is a harmonic two-form with respect to the metric $g$ of $X$ that spans a ``chamber" between two such walls.} 
\be
\Phi_{c_1,k}(p,\{\bfx_i \}) :=  \int_{\CM_k} \mu(\bfx_1) \cup \ldots \cup \mu(\bfx_m) \cup \mu(p)^t,
\ee
and this integral yields a non-zero value only for $s= 2m + 4t$. The information of Donaldson invariants can be repackaged conveniently in a generating function. The Donaldson series is defined, for a formal variable $q$, as
\be
\Phi_{c_1}(e^{\alpha p + \beta \bfx} ) := \sum_{k= 0}^{\infty} q^{k}\sum_{t,m}\Phi_{c_1,k}\left(\frac{p^t}{t!}, \frac{\bfx^m}{m!}\right)\alpha^t \beta^m  = \sum_{k= 0}^{\infty} q^{k}\int_{\CM_{k}} e^{\mu(\alpha p + \beta \bfx)}.
\ee
 If the underlying manifold is an algebraic surface then the moduli spaces of instantons can be identified with the (compactified) moduli space of  $\mu$-stable vector bundles, or better, coherent torsion free sheaves, and the integrals yielding Donaldson invariants can equivalently be defined over these moduli spaces. For an introduction see \cite{DONALDSON1990257,Donaldson90, MooreNotes2017} but also see the monograph \cite{Mochizuki} for a thourough analysis of the algebro-geometric analogues of Donaldson invariants that have lead to some new conjectures \cite{Gottsche:2017vxs}. 

Donaldson theory turns out to have a physical (or rather, supersymmetric) counterpart, the so-called Donaldson-Witten theory, which is the topologically twisted $\CN=2$ super Yang-Mills theory on a smooth four-manifold $X$ with gauge group $\mathrm{SU}(2)$ as it was shown in \cite{Witten:1988ze}. When this theory is evaluated on a four-manifold $X$, correlators of some specific operators reproduce the Donaldson invariants of the four-manifold. This quite remarkable result together with the equally important solution of the low energy effective theory of the $\CN=2$ SYM theory on $\mathbb{R}^4$, the Seiberg-Witten theory \cite{Seiberg:1994aj,Seiberg:1994rs}, as well as the definition of the Seiberg-Witten invariants \cite{Witten:1995gf} led eventually to the work of Moore and Witten \cite{Moore:1997pc}. In that work the physics of twisted $\CN=2$ theories on arbitrary four-manifolds was determined in terms of the so-called $u$-plane integral which we will briefly recall in Section \ref{uplaneOLD}. For an introduction see \cite{Laba05, MooreNotes2017}. The $u$-plane integral, or Coulomb branch integral, has been somewhat revitalized recently with the works \cite{Korpas:2017qdo,Moore:2017cmm, Korpas:2018} where using the $u$-plane technology some new connections with the theory of indefinite theta functions, mock-modular forms and with non-Lagrangian theories have been made. 

In this paper we continue towards this paradigm by further investigating the ramified $u$-plane integral inspired by the works of Tan \cite{Tan:2009qq,Tan:2010dk}. We study Donaldson-Witten theory in the presence of surface defects that support supersymmetric surface operators. Mathematically these defects correspond to real codimension two surfaces embedded (not necessarily trivially) in $X$. In the algebraic setting these surfaces are genus $g$ smooth algebraic curves. In the presence of such defects it is possible to define the so-called \emph{ramified Donaldson invariants} associated to $X$ \cite{kronheimer1995}.  In the quantum field theory side defects correspond to specific non-local operators and they are not something remarkably new. A familiar (albeit slightly different) example of one-dimensional defects (codimension three with respect to $X$) are line operators like the electrically charged Wilson lines and the magnetically charged t' Hooft (disorder) operators. Our interest will be focused on two-dimensional defects. The study of such operators in supersymmetric quantum field theories was initiated around ten years ago with the works  \cite{Gukov:2006jk, Gukov:2008sn, Gukov:2007ck} for $\CN=4$ theories,  \cite{Gukov:2007ck, Alday:2009fs, Tan:2009qq, Tan:2010dk} for $\CN=2$ theories, \cite{Koh:2009cj} in the context of Klebanov-Witten theory and very interestingly \cite{Cirafici:2013tna} in the context of higher dimensional cohomological field theories\footnote{By CohFT we mean a field theory with a nilpotent scalar symmetry $\mathscr{Q}$ whose observables belong to the $\mathscr{Q}$-cohomology. With this definition Donaldson-Witten theory, as well as other twisted $\CN=2$ theories are CohFTs. Note that this is not the definition of CohFTs that mathematicians use.} (CohFTs) and Donaldson-Thomas theory, aspects of which we hope to return to in the future. A general and complete treatment of surface operators (mainly for $\CN=2$ theories) is found in \cite{Gukov:2014gja} along with many references within.  In the current work take a fresh look at the ramified $u$-plane integral by adding to the Lagrangian of the ramified Donaldson-Witten theory a $\overline{\CQ}$-exact surface operator that couples to the self-dual part of the curvature of the (in an appropriate sense \emph{extended}) gauge bundle, in the presence of surface defects. Such type of operator has previously been considered in \cite{Moore:1998et} in the context of interpreting Witten-like indices in type IIA string theory as correlators of CohFT. This $\overline{\CQ}$-exact insertion allows us to write the $u$-plane integral as an integral of a total anti-holomorphic derivative of an indefinite theta function \`{a} la Zwegers \cite{ZwegersThesis}.

 Indefinite theta functions are examples of mock modular forms and they have appeared in various contexts in the theoretical physics literature. Some examples are found in Vafa-Witten theory \cite{Vafa:1994tf}, in conformal field theory \cite{EguchiTao, Semikhatov:2003uc, Troost:2010ud}, AdS$_3$ gravity \cite{Manschot:2007ha}, supersymmetric black holes \cite{Manschot:2009, Dabholkar:2012nd, Alexandrov:2016tnf} and also in the umbral and Mathieu moonshine phenomenon \cite{Cheng:2011ay}, while a particular type of mock modular forms called skew holomorphic Jacobi forms appeared very recently in the context of half-BPS state counting for heteroting strings compactified on $\mathbb{S}^1$ \cite{Cheng2017}. 
 
 Schematically, an indefinite theta function is a map $\Theta : \BH \to \BC$ which can be expressed as a holomorphic $q$-series over a lattice $\Lambda$ of indefinite signature (we will be considering Lorentzian lattices in this paper but see \cite{Manschot:2017xcr} for an example where a lattice of signature $(2,n-2)$ appears in the context of $\mathrm{SU}(3)$ Vafa-Witten theory). As usual, we define $q = e^{2\pi i \tau}$. The $q$-series $\Theta$ converges since we only consider vectors $\bfv \in \Lambda$ such that they have negative definite norm (following the conventions of \cite{Korpas:2017qdo}). Such functions fail to be modular under $\mathrm{SL}(2,\BZ)$ though and the way to go around this is to add a very specific non-holomorphic function $R$, whose argument is a function $g: \BH \times \bar{\BH}\to \BC$,  such that the sum $\widehat{\Theta}=\Theta + R$ transforms as a modular form (although it is not a holomorphic function anymore). The argument $g$ of $R$ is called \emph{the shadow\footnote{The term ``shadow of a mock modular form" first appears in the work of Zagier \cite{DonZagier}.} of} $\widehat{\Theta}$. If we take the anti-holomorphic derivative of $\widehat{\Theta}$ we find that it is equal to a Siegel-Narain theta function defined on the same lattice $\Lambda$ (which is associated to the path integral of the gauge field of the theory under consideration as we will see). As we just explained, this procedure is achieved in the setting of (ramified or not) Donaldson-Witten theory exactly by the inclusion of such a $\overline{\CQ}$-exact surface operator which does not modify the theory as by the standard rules of topological field theories \cite{Witten:1988ze, WittenCohTQFT}, a non-trivial statement studied in detail \cite{Korpas:2018} where it is shown that this $\overline{\CQ}$-exact operator is a well defined observable of the low energy physics. The $u$-plane integral then localizes to the cusps of the integration domain which for the case of pure theory (without any matter representations) they are the points $\{0,2, i\infty\}$ of $\BH/\Gamma^{0}(4)$. If the four-manifold admits a metric of positive scalar curvature then the only contribution comes from the cusp at infinity (since for such class of manifolds the Seiberg-Witten invariants vanish) and the result of the integral is given in terms of the $q^0$ coefficient of the holomorphic theta function. This property of the $u$-plane integral has also been recently discussed in \cite{Korpas:2017qdo} and it is a hope that it will be able to shed new light to the $u$-plane integral for the conformal theories $N_f=4$ and mass deformed $\CN=4$ as well as for the topologically twisted class-$\CS$ theories, the latter of which are candidates of theories that can provide an alternate direction in the search for new smooth four-manifold invariants \cite{Moore:2017cmm, Dedushenko:2017tdw, Gukov:2017zao} . Although our considerations do not change dramatically the scenery in the world of four-manifold invariants\footnote{It is known from the works of Kronheimer and Mrowka that the ramified Donaldson invariants do not provide new smooth structure invariants compared to the usual Donaldson invariants.} we do show that the $u$-plane integral in terms of indefinite theta functions is consistent with the theories defined on four-manifolds on the presence of surface defects and therefore the possibility of including such operators in the previously mentioned theories is open. 

\subsection*{An appetizer}
The literature on surface operators (as well as line operators and domain walls) is vast, nevertheless, we would like to begin with a simple appetizer to get some intuition.

We will see how line operators (co-dimension three) can arize in classical gauge theory. For detailed exposition in these lines see \cite{Deligne99quantumfields}. Line operators arise from some very simple considerations in classical field theory. Consider a vector bundle $\CE \to X$ with connection $A$ over a $C^{\infty}$ manifold $X$ and structure group a simple Lie group $G$. The Lagrangian (density) of the theory is 
\be
\mathscr{L}(A) = -\frac{1}{2} F_A \wedge *F_A
\ee
where the trace over the Killing form is implied and with $F_A = dA + [A,A]$ as usual. In principle we can couple the theory to an external current $J \in \Omega^{3}(X)$ that is assumed to be conserved, i.e.,  $dJ = 0$ and therefore we can write $J = dK$. This would give a modified Lagrangian, namely
\be 
\mathscr{L}(A) = -\frac{1}{2} F_A \wedge *F_A + J \wedge A
\ee 
and the equations of motions would include sources. Generically, this expression is ill-defined because the second summand is not gauge invariant. Nevertheless, since we required $J=dK$ we can invoke Stokes' theorem and write 
\be
\int J \wedge A = - \int K \wedge F_A
\ee
which is a well-defined expression provided there is a suitable decay at infinity. The interpretation is that this current could be induced by a point-like particle with electric charge $q$ transversing  a trajectory $s$ through $X$. The current $J$ then is Poincar\'e dual to  homology class $s$. 

\begin{center}
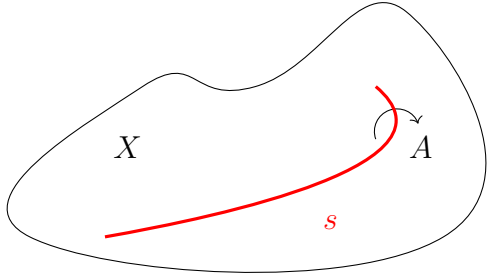

 \begin{tikzpicture}
    \draw[smooth cycle,tension=.9] plot coordinates{(-1,0) (0.5,2) (2,2) (4,3) (4.5,0)};
    \coordinate (A) at (3.6,1.3);
    \draw[->] (A) arc(200:20.0:0.3) ;
    \draw[very thick, red] (0,0) to[out=10,in=-40] (3.6,2.0);
    \node (C) at (3.0, 0.2) {\color{red} $s$};
     \node (C) at (0.3, 1.2) { $X$};
     \node (C) at (4.2, 1.2) { $A$};
  \end{tikzpicture}
  \captionof{figure}{ An embedded curve (red) $s$ in a smooth manifold $X$. The connection in the neighborhood of $s$ has a particular behavior in the monodromy. } 
\end{center}

The curve $s$, viewed now as an embedded curve in $X$ corresponds to an external current. The gauge field has a particular behavior along a monodromy around $s$. Actually, to define such a line operator it suffices to define a prescribed singular behaviour of the connection $A$ in the vicinity of $s$.  The physics of supersymmetric surface operators  that we will discuss is qualitatively very similar to what we just explained in the sense that these operators have support precisely on embedded surfaces of codimension two.

In topological quantum field theories when we try to compute the partition function of the theory,
\be
{Z}_{\text{Maxwell}} = \int [\mathscr{D}A] e^{-\int_X \mathscr{L}(A) },
\ee
we end up summing up vectors over a particular lattice $\Lambda$ \cite{Witten:1995gf, Verlinde:1995mz}. We will indeed end up summing vectors over such lattice in the ramified Donaldson-Witten theory. The novelty of this paper will be shown to be to associate to such a path integral an indefinite theta function that also depends on an embedded surface, which as described previously is an object which has its roots in analytic number theory \cite{ZwegersThesis}. Due to this property the computation the path integral of the theory, or to be more accurate a specific correlator, will be given by a quite simple formula.  \\

The structure of the paper is as follows. In Section \ref{Ramified Donaldson invariants} we review the notion of ramified Donaldson invariants. In Section \ref{Sec2} we recall how supersymmetric surface operators appear in $\CN=4$ and mainly in $\CN=2$ theories in four dimensions. In Section \ref{uplaneOLD} we give a quick overview of the $u$-plane integral as it appears in the usual Donaldson-Witten theory. In Section \ref{TheModiefied} we describe in some detail how the $u$-plane is modified in the presence of the embedded surfaces, we include the $\bar{\CQ}$-exact deformation and show that the integral localizes at the cusp at infinity for specific manifolds of Kodaira dimension $-\infty$. In Section \ref{LAST} we conclude with some remarks and some discussion. In Appendix \ref{App1} we discuss in some detail how to surface operators correspond to lifts of the maximal torus to the Cartan subalgebra. In Appendix \ref{App0} we discuss the modularity properties of the Siegel-Narain theta function that appears in the main text and in Appendix \ref{AppA} we define Zwegers' indefinite theta functions and highlight some of their properties.


\section{Ramified Donaldson invariants}  \label{Ramified Donaldson invariants}
In the introduction we gave a brief review of the definition of the usual Donaldson invariants (without embedded surfaces). In this section we will briefly review the notion of ramified Donaldson invariants following closely \cite{kronheimer1995, math/9404232,Tan:2009qq}. Let $X$ be a smooth, closed (compact without boundary) and simply connected four-manifold equipped with a Riemannian metric. Let $\CE$ be a $G= \SOT$ principal bundle over $X$ that can be lifted to a $G=\SU$ principal bundle for $w_2(\CE)=0$. We denote by $\mathsf{g} = \mathrm{Lie}(G)$ and by $\mathsf{t} = \mathrm{Lie}(\mathbb{T})$ the Cartan subalgebra, where $\mathbb{T}$ is the maximal torus of $G$. Note that the middle integral homology is isomorphic to a lattice $\Lambda = \BZ^{b_2}$ and splits as two orthogonal components $\BZ^{b_2^+,0} \perp \BZ^{0,b_2^{-}}$ where we have exactly $b_2=b_2^{+}+ b_2^{-}$. The lattice comes equipped with a unimodular quadratic form $Q:H^2(X, \BR)\to \BR$ and a bilinear form $B: H^2(X, \BR) \times H^2(X, \BR) \to \BR$. Explicitly they read
\be \label{QUADBIN}
\begin{split}
Q(\bfk) & := \int_X \bfk \wedge \bfk : = \bfk^2, \\[0.5em]
B(\bfk, \bfl) &:= \int_X \bfk \wedge \bfl, \ 
\end{split}
\ee
for $\bfk, \bfl \in \Lambda$. By restricting only to integeral classes in $H^2(X,\BZ)$ both the quadratic and bilinear forms are $\BZ$ valued.  An embedded closed surface $\boldsymbol{S} \hookrightarrow X$ is a genus $g$ complex curve\footnote{The genus of this curve can be determined by the adjunction formula
$ g(\bfS) = 1 + \frac{1}{2}(\bfS^2 + B(\bfS , K_{X}))$.} embedded into $X$. Therefore by $Q(\bfS)$ we denote the self-intersection number of $\bfS$  and also by $\bar{X} = X \backslash \bfS$ we denote the complement of $\bfS$. Near an open neighborhood of $\bfS$ we can split the vector bundle to a sum of complex line bundles over $X$ as  $\CE = \CL \oplus \CL^{-1}$. We denote the curvature of a connection on $\CL$ by $F_{\CL}$.  Let $A \in \Omega^1(\bar{X}, \mathsf{g})$ denote the  local $G$-connection one-form which becomes singular as it approaches $\bfS$. Then locally this connection takes the form 
\be 
A = i\alpha d\theta + \mathrm{regular}
\ee
with $\alpha \in \sigma_3 \BR$ and $\sigma_3 = \mathrm{diag}(1,-1) \in \mathsf{t}$. The angular variable comes from $z = r e^{i\theta}$ where $z$ is a holomorphic coordinate normal to $\bfS$ and the connection is singular as $z \to 0$. Due to the coordinate singularity we just discussed there exists a non-trivial gauge-invariant holonomy $\text{Hol}_{\gamma}(A) = e^{-2\pi \alpha}$ of the connection contouring some small loop $\gamma$ around $\bfS$.  Note though that if $\text{Hol}_{\gamma}(A)$ is trivial, such that $\gamma$ is contractible, then the connection is an ordinary connection on $\CE$ defined over $X$.  These embedded surfaces will turn out to support \emph{supersymmetric surface operators}. 

Using the above we can define the so-called \emph{ramified Donaldson invariants} for a four-manifold $X$ with the presence of an embedded complex curve $\bfS$. The ramified Donaldson invariants $\cD$ are defined very similarly to the ordinary ones, that is they are polynomials on the homology of $\bar{X}$ with rational coefficients
\be
\cD : \mathrm{Sym}[H_0(\bar{X},\BQ) \oplus H_2(\bar{X},\BQ)] \to \BQ.
\ee
Let ${\CM}_{{\tilde{k}}}$ denote the moduli space of ramified $G$-instantons with instanton number ${\tilde{k}}= \int_{\bar{X}} p_1(\CE)/4$ where $p_1(\CE)$ is the first Pontryagin class of $\CE$. Also recall that the Euler characteristic of $X$ is $\chi = \sum_i(-1)^ib_i$, the signature of $X$ is $\sigma=b_2^{+}-b_2^{-}$. By $l = -\int_{\bfS} c_1(\CL)$ we denote the magnetic flux number.

The dimension of the moduli space of ramified $G$-instantons, that for brevity we denote by $s:=\mathrm{dim}({\mathcal{M}}_{\tilde{k}})$, is 
\[
s=8k-\frac{3}{2}(\chi+\sigma)+4l-2(g-1)
\]
where $k = -\int_X ch_2(\CE)$ and we assume no reducible connections. Then, the corresponding degree $s$ ramified Donaldson invariants are defined as  
\be
\cD_{w_2}^J(p,\bfx) = \sum_{2m+4t=s} p^t \bfx^m P_{w_2,\tilde{k}}.
\ee
where $J$ denotes a choice of polarization in $H_2(X,\BR)$. The ramified Donaldson invariants are the numbers $P_{w_2,\tilde{k}}$ and are given, like in the case of ordinary Donaldson invariants, through the intersection theory of ${\mathcal{M}}_{{\tilde{k}}}$. In analogy to the ordinary case, there exists a universal ramified instanton bundle and a ramified slant product (or Donaldson map) $\bar{\mu}_D$ such that 
\begin{equation}
\bar{\mu}_D : H_i(\bar{X}, \BQ) \to H^{4-i}({\mathcal{M}}_{\tilde{k}}, \BQ).
\end{equation}
Therefore, for two homology classes $p \in H_0(\bar{X},\BQ)$ and $\bfx \in H_2(\bar{X},\BQ)$ we have 
\be
P_{w_2,\tilde{k}}(p^t, \bfx^m) = \sum_{r,s \geq 0} \int_{{\mathcal{M}}_{\tilde{k}}} \mu_D(\bfx)^m \cup \mu_D(p)^t  .
\ee
We can write down the generating function of ramified Donaldson invariants, which physically corresponds to a specific correlation function (that we discuss in detail in Section \ref{TheModiefied}), by summing over all vector bundles $\CE \to X$ for a fixed $w_2(\CE)$ and varying $\tilde{k}$. Therefore the generating function of ramified Donaldson invariants is defined as
\begin{eqnarray}
\tilde{\Phi}_{\bfmu}^J(p, \bfx)&=&  \sum_{\tilde{k}=0}^{\infty}q^k P_{w_2, \tilde{k}}(e^p, e^{\bfx}) \\
												&=&  \sum_{\tilde{k}=0}^{\infty} q^k \sum_{t,m \in \BZ_{\geq 0}} \frac{p^t}{t^!}\frac{\bfx^m}{m!}P_{w_2,\tilde{k}}(p^t, \bfx^m), 
\end{eqnarray}
where $\bfmu \in c_1(\CE)/2 + H^2(\bar{X}, \BZ)$. Following \cite{Moore:1997pc, Korpas:2017qdo} we refer to $\tilde{\Phi}_{\bfmu}^J(p, \bfx)$ as the \emph{partition function}. Just like in the case of ordinary Donaldson invariants, if $b_2^{+}(X)=1$ then the partition function develops a metric dependence and wall-crossing phenomena, that is $\tilde{\Phi}_{\bfmu}(p, \bfx)$ jumps discontinuously when crossing a wall of marginal stability in the positive cone\footnote{See subsection \ref{WCformula} for definition.} of $X$ and in general we have $\tilde{\Phi}_{\bfmu}^{J}(p, \bfx) \neq \tilde{\Phi}_{\bfmu}^{J'}(p, \bfx)$ for two different period points $J,J' \in H^2(X,\BR)$. Therefore, strictly speaking, these partition functions are not quite smooth structure invariants, rather piecewise invariants. The difference of the Donaldson polynomials between two period points belonging to different chambers is given by the wall-crossing formula that will be discussed in Section \ref{TheModiefied}.


\section{Review of surface operators in four dimensions} \label{Sec2}
In this section we will briefly recall some well known facts about surface operators in $\CN=2$ supersymmetric Yang-Mills theories. Surface operators in topological field theories first appeared in \cite{Gukov:2006jk} where the authors consider the GL-twist of the $\CN=4$  super Yang-Mills theory, the Kapustin-Witten topological field theory \cite{Kapustin:2006pk}. In order to set the stage and get some intuition about surface operators we begin by considering the non-compact four-manifold $\BC^2 \cong \BC_a \times \BC_b$ spanning $x_0, x_1, x_2, x_3$ and consider a codimension two defect supported at  $\BC_a$ with coordinates $x_0$ and $x_1$ which is localized at $x_2=x_3=0$. 

\begin{figure}[h]
\includegraphics[width=8cm]{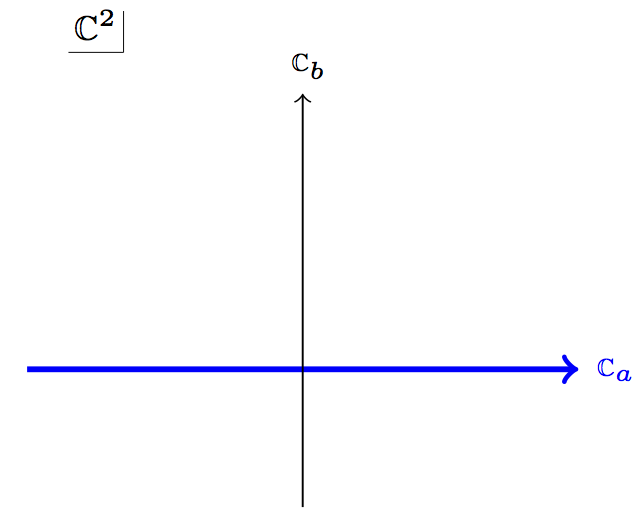}
\centering
\label{SurfaceDefect}
\caption{A surface defect supported on $\BC_a \subset \BC^2$.}
\end{figure}

The bosonic fields on $\BC_a$ are the two components of the original gauge field spanning $\BC_a$, that is $A_0,A_1$ and four of the six scalars of the $\CN=4$ multiplet. There is a two dimensional supersymmetric theory living on $\BC_a$. Along $\BC_b$ it is required that the normal components $A_2,A_3$ for the gauge field and $\phi_2, \phi_3$ for the scalars, have a suitable singular behaviour as they approach $\BC_a$. Supersymmetry then requires that $A = A_2dx^2 + A_3dx^3$, $F=dA$ and $\phi = \phi_2 dx^2 + \phi_3 dx^3$ satisfy Hitchin's equations
\begin{eqnarray}
F + \phi \wedge \phi &=& 0, \\
d_A \phi &=& 0, \\
d_A * \phi &=& 0. \
\end{eqnarray}
Next we want to set $x_2+ix_3 =z = re^{i\theta}$ and move to $\CN=2$ supersymmetry where it is no more true that we have six scalar fields. Theories with $\CN=2$ supersymmetry in four dimensions admit half-BPS surface operators and the corresponding two-dimensional theory on $\BC_a$ preserves $\CN=(2,2)$ supersymmetry \cite{Gaiotto:2009fs}. The BPS equations for the (untwisted) $\CN=2$ theory are $F=0$ and $d_A\phi =0$, where $\phi$ is the adjoint valued Higgs field. Let us consider the simple case where $\phi=0$. Then the BPS configurations correspond to irreducible flat connections on $X$. Therefore any such surface operator is in a one-to-one correspondance with an irreducible flat connection on the vector bundle $\CE \to X$ restricted on $\bar{X}$ which is singular along $\bfS$ or, in other words, a surface operator supported along $\bfS \hookrightarrow X$ corresponds to an irreducible  flat connection on $\CE \to \bar{X}$. The flatness condition is obvious by taking Hitchin's equations and letting $\phi$ to be trivial. The curvature of the connection should be vanishing on $\bar{X}$ but not on $\bfS$. Instead, we can consider an \emph{extended} vector bundle $\mathscr{E} \to X$ with connection one-form $\mathscr{A}$ such that the curvature two-form is given by $\mathscr{F} = F - 2\pi \alpha \delta_{\bfS}$ with $\alpha$ interpreted as the electric charge of the surface operator. This bundle extension was first implemented in the context of the ramified geometric Langland's program in \cite{Gukov:2006jk} where connection to parabolic Higgs bundles was described. The BPS condition\footnote{Note that for the twisted theory the BPS equation is $\mathscr{F}^{+}=0$ \cite[chapter 5]{Laba05}.} gives the flatness condition 
\be
\mathscr{F}  = 0
\ee
or equivalently $F = 2\pi \alpha \delta_{\bfS}$. Since $F$ is the two-form curvature we can split it in self-dual $F^{+} = 2\pi \alpha \delta_{\bfS}^{+}$ and anti-self-dual $F^{-} = 2\pi \alpha \delta_{\bfS}^{-}$ parts where $\delta_{\bfS}$ denotes the Poincar\'e two-form dual to the homology class $\bfS$ supported on. Let us make a clarification of the notation we use. Note that $\bfS$ belongs to fundamental class of its own homology $H_2(\bfS)$  while we denote by $\delta_{\bfS}$ the two-form Poincar\'e dual of the Dirac delta function supported on $\bfS$. As explained in detail in \cite{Gukov:2006jk, Tan:2009qq,Tan:2010dk} in principle  $\alpha \in \mathbb{T} = \mathsf{t}/\Lambda_{\mathrm{cochar.}}$ where $\Lambda_{\mathrm{cochar.}} = \text{Hom}({\rm U}(1),\mathbb{T})$ is the co-character lattice \cite{Gukov:2006jk}. This is not quite the case though because we have extended the bundle $\CE$ to $\mathscr{E}$ and this amounts in lifting $\alpha$ to the Cartan subalgebra $\mathsf{t}$ and there are many inequivalent such lifts yielding the same holonomy for $A$. Each possible lift corresponds to a different surface operator therefore. We include some discussion on the lifts of surface operators in Appendix \ref{App1}. Note, that in this extended bundle $\mathscr{E}$ we can include a ``theta-like angle" term with a contribution to the path integral as 
\be
e^{i \eta \int_{\bfS} F },
\ee
where the exponent measures the magnetic flux of $\mathscr{E}$ through $\bfS$. Note that this integral is proportional to the monopole number, with $\eta \in \mathsf{t}$ as well. In that sense the pair $(\alpha,\eta)$ corresponds to electric and magnetic charges of a dyon-like surface operator that we can interpret to be supported on $\bfS$. Mathematically such type of extensions of vector bundles can be described in the context of parabolic bundles \cite{Mehta1980, Cirafici:2013tna} which are very interesting objects on their own with connections to the Riemann-Hilbert problems or the Painvel\'e equations among others (for example see \cite{article} for parabolic bundles over $\BC \BP^1$). We hope to expand in this context in a future work on ramified Vafa-Witten theory (that is the theory in the presence of embedded divisors).




\section{The $u$-plane integral} \label{uplaneOLD}
The classical moduli space of vacua of $\CN=2$ theories is $\mathsf{t}/W$ where $W$ is the Weyl group associated to $G$. For for rank one gauge algebras the coordinate ring of $\mathsf{t}/W$ is identified with the equivariant cohomology of a point $H_{G}^{*}(pt)$ which has a single generator $u := \frac{1}{16\pi^2}{\rm Tr}(\phi^2)$. The Coulomb branch $\CB$ is a quantum lift of the classical moduli space of vacua defined through the moduli space of instantons \cite{Moore:1997pc, Nakajima:2003uh}. The natural coordinate is  again $u$ that now belongs to the field of fractions of $H_{G}^{*}(pt)$. In Donaldson-Witten theory the partition function $ {Z}_{\text{DW}}$, receives contributions from two different terms:
\be \label{UplusSW}
{Z}_{\text{DW}} =  {Z}_u + {Z}_{\text{SW}}.
\ee 
The left hand side of the equality above corresponds precisely to the generating function of (ramified) Donaldson invariants. The right hand side is composed by $Z_u$, the $u$-plane integral, that we will explain in detail below, and ${Z}_{\text{SW}}$ given by
\be 
{Z}_{\text{SW}} = \sum_{s} {Z}(u_s),
\ee
where we sum contributions to ${Z}_{\rm DW}$ from all $s \in \CB$ such that the discriminant of the Seiberg-Witten curve is zero, i.e., the Seiberg-Witten contributions. That is, the union of $s$ is a divisor along which the elliptic fiber over $\CB$, which is the Seiberg-Witten curve
\[
y^2 = 4x(x^2 - ux + \frac{1}{4}\Lambda^4),
\]
becomes singular. Here $x,y \in \BC$ and $\Lambda$ is usually called ``the dynamically generated mass scale" of the theory and we can take it to be one. Note that this is a curve with respect to the $\Gamma^0(4) \subset \mathrm{SL}(2,\BZ)$ congruence subgroup\footnote{The subgroup $\Gamma^0(n)$ is defined as the subgroup  matrices of the form $ \left( {\begin{array}{cc}
   a & b \\
   c & d \\
  \end{array} } \right) \in \mathrm{SL}(2,\BZ)$ such that $b = 0 \text{ mod } n$.}. For four-manifolds with $b_2^+=1$ that admit a metric of positive scalar curvature these contributions ${{Z}}_{\rm SW}$ vanish \cite{Moore:1997pc} (see Section \ref{Anote}) but our result is independent of this fact. 

\subsection{A note about four-manifolds} \label{Anote}

Let us stress for another time that the $u$-plane integral only contributes for manifolds $X$ with $b_2^{+}(X) = 1$. To be more precise, manifolds with $b_2^+=0$ also allow the study of the $u$-plane integral but for such manifolds the calculations are more involved due to the presence of one-loop determinants. We will be focusing on simply connected four-manifolds with $b_2^{+}(X) = 1$ then (which by a theorem of Wu are always at least almost complex manifolds).  Let us note that they are somewhat special in the extraordinary world of four-manifolds due to the wall-crossing phenomena that appear in their Donaldson invariants as observed by G\"ottsche and Zagier \cite{Gottsche:1996aoa}.  Some ``easy to work with" examples of manifolds with $b_{2}^+=1$ are K\"ahler surfaces of Kodaira dimension $-\infty$, i.e. ${\rm dim}(H^0(X, K_X)) = 0$, where we denote by $K_X := c_1(\CK_X)$, the first Chern class of the canonical line bundle $\CK_X \in \mathrm{Pic}(X)$ \cite{Barth, griffiths2011principles}. These K\"ahler surfaces come in three families. Let $X$ be a K\"ahler surface of such type. 
\begin{enumerate}
\item  If $K_X^2>0$ the surface $X$ is rational or ruled,
\item  if $K_X^2=0$ then the surface is a $\BC \BP^1$ bundle over $\mathbb{T}^2$,
\item  if $K_X^2<0$ then the surface is a $\BC \BP^1$ bundle over a curve $C_g$ of genus $g$ greater than one.
\end{enumerate}
For the first case, the rational and  ruled surfaces, the Seiberg-Witten contribution vanishes exactly because they admit a K\"ahler metric of positive scalar curvature. Specific examples of surfaces that have $b_2^+=1$ and positive scalar curvature are: the projective plane $\BC \BP^2$, del Pezzo surfaces (blow-ups of the projective plane up to nice points), Hirzebruch surfaces $\BF_l$ (they are defined as the projectivizations of the bundle $\CO_{\BC \BP^1} \oplus \CO_{\BC \BP^1}(-l)$), see. \cite{Barth}. As a matter of fact, it is a theorem that if a K\"ahler surface $X$ admits a metric of positive scalar curvature then it is rational or ruled \cite{LeBrun1995}.  We want to stress the importance of such surfaces due to the fact that they allow us to probe the Coulomb branch of the theory (in both the usual and ramified versions) while most four-manifolds will not allow for this. These four-manifolds, therefore, provide an excellent lab to study quantitatively and qualitatively the low energy effective theories in their totality.

\subsection{Remarks on the $u$-plane integral} \label{FirstLook}
The first summand in (\ref{UplusSW}) therefore is called $u$-plane integral and it amounts to the full the path integral of the low energy effective Donaldson-Witten theory when the Seiberg-Witten contributions vanish. A more detailed discussion of the derivation of the  $u$-plane integral for the ramified theory will be given in Section \ref{TheModiefied}, but let us first recall some facts about the integral for the usual Donaldson-Witten theory and discuss some of its features. The $u$-plane integral was first derived in \cite{Moore:1997pc} and takes the form
\be \label{Zu}
{Z}_u(p,\bfx) = \int_{\CB} da \wedge d\bar{a}~ A(u)^{\chi} B(u)^{\sigma} e^{2pu + \bfx^2G(u)} \Psi_{\bfmu}^J(\tau, \bfrho).
\ee
Let us explain the terms and the notation we see in (\ref{Zu}) in order to get a better understanding of what this integral actually is. 
\begin{itemize}
\item $\CB$, as mentioned earlier, denotes the Coulomb branch of the low energy effective Donaldson-Witten theory. It is the so-called $u$-plane and it is isomorphic to $\BH/\Gamma^{0}(4)$ which has three cusps at 0 (monopole point), 2 (dyon point), and $i\infty$ (the semi-classical limit).   
\item $a$ is the Coulomb branch special coordinate that arises from the Higgs field of the $\CN=2$ vector multiplet and corresponds to the period of the Seiberg-Witten differential $\lambda_{\text{SW}}$ on a cycle of the Seiberg-Witten curve under the path $u \to e^{2\pi i} u$ near $u \to \infty$. Its \emph{magnetic} dual is $a_D$ and it is given by integrating the Seiberg-Witten differential over the other cycle of the Seiberg-Witten curve. For a theory of rank $r$ there are $r$ of each.
\item $A$ and $B$ are terms that depend on the parameters of the $\CN=2$ theory and the Riemannian metric on $X$ while $\chi$ is the Euler character of $X$ and $\sigma$ is the signature of $X$. These terms arise due to the necessity to include gravitational coupling to the theory when we want to evaluate it in a closed smooth four-manifold \cite{Witten:1995gf, Moore:1997pc} (we expand on this point in Section \ref{GravCoup} of this paper).  Specifically, for the Donaldson-Witten theory (also its ramified version)
\begin{eqnarray}
A(u) &=& \alpha \left( \frac{du}{da} \right)^{\frac{1}{2}} \\[0.6em]
B(u) &=& \beta \Delta^{\frac{1}{8}} \
\end{eqnarray}
with $\alpha, \beta$ functions of the parameters of the theory and independent of $X$. Also, $\Delta$ is the discriminant of the Seiberg-Witten curve that determines the low energy theory. The terms $A$ and $B$ combine to the so-called \emph{measure term} $\nu(\tau)$ that will be of importance in Section \ref{TheModiefied}. The similarity of these terms to ``graviational couplings" of two-dimensional theories obtained by topological reduction of non-Lagrangian SCFTs have recently been discussed in \cite{Gukov:2017zao}.

\item The integral depends on two classes $p \in H_0(X)$, $\bfx \in H_2(X)$ while $G(u)$ is called the \emph{contact term} which is crucial for preserving modular invariance of the integrand. This operator is a result of the IR flow of the theory. A detailed analysis of such terms appear in \cite{Moore:1997pc, LoNeSha, Laba05}.
\item $\Psi_{\bfmu}^J(p,\bfx)$ is a Siegel-Narain theta function that captures the contributions to path integral from the ${\rm U}(1)$ Maxwell sector of the theory. When considering such a theory on a four-manifold we can factorize the photon path integral into a term arising from the classical saddle points and a term proportional to the ratio of the determinants of the kinetic operators of the ghost fields and the gauge field. It is precisely the contribution of the saddle points that is responsible for the appearence of the theta function. A detailed analysis of such terms is found in \cite{Witten:1995gf, Moore:1997pc}.
\end{itemize}
In the presence of surface defects the $u$-plane integral will be altered slightly, mainly due to the fact that we consider a different (extended) vector bundle than the one considered in \cite{Moore:1997pc, Tan:2009qq, Korpas:2017qdo}. A detailed derivation of the $u$-plane integral can be found in \cite{Moore:1997pc, Laba05} while for the notation we use in this paper see \cite{Korpas:2017qdo}.


\section{The ramified $u$-plane integral}\label{TheModiefied}
The ramified $u$-plane integral is the path integral over the Coulomb branch $\CB$ of the Donaldson-Witten gauge theory in the presence of surface defects with gauge group $\mathrm{SU}(2)$ or  $\mathrm{SO}(3)$. For explicit details of the derivation one can consult \cite{Moore:1997pc} and \cite{Tan:2009qq}. Let us first, write down the Lagrangian of the Donaldson-Witten theory in the presence of surface defects. 
\be
\begin{split}
\mathscr{L} &= \frac{i}{16\pi} (\bar{\tau}|\mathscr{F}_{+}|^2 + \tau |\mathscr{F}_{-}|^2 ) + \frac{\mathrm{Im}\tau}{8\pi}da \wedge *d\bar{a} -   \frac{\mathrm{Im}\tau}{8\pi} D \wedge *D - \frac{\tau}{16} \psi \wedge *d\eta + \frac{\bar{\tau}}{16\pi} \eta \wedge * d\eta \\[6pt]
                   & \quad +  \frac{\tau}{8\pi} \psi \wedge d\chi - \frac{\bar{\tau}}{8\pi} \chi \wedge d\psi + \frac{\sqrt{2}i}{16\pi}\frac{d\bar{\tau}}{d\bar{a}} \eta \chi \wedge (\mathscr{F}_{+}+ D) - \frac{\sqrt{2}i}{2^7\pi} \frac{d\tau}{da}(\psi \wedge \psi)\wedge(\mathscr{F}_{-}+ D) \\[6pt]
                   & \quad + \frac{i}{4}\eta_{\mathrm{m.}} \mathscr{F}\wedge \delta_{\bfS} + \frac{i}{2^{11}\cdot 3\pi} \frac{d^2\tau}{da^2} \psi \wedge \psi \wedge \psi \wedge \psi - \frac{\sqrt{2}i}{3 \cdot 2^5 \pi} \{\overline{\CQ}, \frac{d\bar{\tau}}{d\bar{a}} \chi_{\mu \nu} \chi^{\nu \lambda}\chi_{\lambda}^{\mu} \}\mathrm{vol}_X.
\end{split}
\ee
This Lagrangian is identical to the one of the usual Donaldson-Witten theory except for the term that is proportional to $\eta_{\mathrm{m}}$ which appears\footnote{There is a clash of notation here. In the Lagrangian by $\eta$ we denote the Grassman valued zero-form of the theory and by $\eta_{\mathrm{m}}$ we denote the ``magnetic charge" of the surface defect that we denoted as $\eta$ earlier. We will eventually integrate out the zero-form $\eta$ and thus we will be able to return to the previous notation for the magnetic charge. } due to the presence of the surface defect and the fact that we consider the field strengths $\mathscr{F}$ for connections on the extended vector bundle $\mathscr{E}$. Recall that Donaldson-Witten theory contains a BRST-like nilpotent scalar supercharge $\overline{\CQ}$ whose cohomology provides the physical observables of the theory.  The supersymmetric algebra in the presence of the surface defect reads
\be
\begin{split}
&[\overline{\CQ}, \mathscr{A}] = \psi \hspace{4em}  [\overline{\CQ}, a] = 0   \hspace{4em}   [\overline{\CQ}, \bar{a}] = \sqrt{2}i\eta \\
&[\overline{\CQ}, D] = d_{\mathscr{A}}\psi  \hspace{2.9em} \{\overline{\CQ}, \psi \} = 4\sqrt{2}da \\
&\{\overline{\CQ}, \eta  \} = 0 \hspace{4.2em}   \{\overline{\CQ}, \chi  \} = i(\mathscr{F}- D)
\end{split}
\ee
with $\mathscr{A}$ the connection of the low energy U(1) line bundle (in the Coulomb branch of the theory). The path integral of the theory is 
\[
Z_\varphi = \int [\mathscr{D}\varphi]~ e^{-\int_X \mathscr{L}[\varphi]},
\]
where by $\varphi$ we collectively denote all the fields of the theory. Nevertheless, this is not quite what we are interested in, rather we need to take into account a few further things. As we briefly discussed previously we need to take into account gravitational coupling whose modular properties ensure that the integral is modular invariant. Furthermore we need to calculate a specific correlation function, not just the partition function, and this correlation function will depend on a point and a surface operator that have a non-trivial IR flow, as well as the embedded surface. In this section we discuss these issues and derive the explicit form of the $u$-plane integral by making a modification to the story of \cite{Tan:2009qq} by adding a $\overline{\CQ}$-exact operator that will make the $u$-plane integral behave in a particularly nice way.

\subsection{The gravitational couplings}\label{GravCoup}
In order the low energy effective Donaldson-Witten theory to be consistent we need to include to the Lagrangian some further couplings to the background curvature of $X$, in specific these terms will be proportional to the Euler number and signature of $X$. Such terms were first derived by Witten in \cite{Witten:1995gf} by $R$-symmetry anomaly arguments. Schematically these terms take the form
\be
 (\log A(u)) \, \mathrm{tr} R \wedge *R + (\log B(u)) \mathrm{tr} \, R\wedge R
\ee
where $A$ and $B$ are holomorphic functions in $u$. These are precisely the terms that we saw in Section \ref{FirstLook}. These terms can be evaluated for a generic $\CN=2$ theory, Lagrangian or non-Lagrangian. For theories with Lagrangian description these terms can be obtained from the calculation of the prepotential $\CF_0(a)$ of the corresponding Seiberg-Witten theory. In specific, the logarithm of the Nekrasov partition function (the free energy) is identified with the Seiberg-Witten prepotential $\CF_0(a)$ at the limit $\epsilon_{i=1,2}=0$.
\[
 \log Z_{\text{Nek.}}(a,\eps_1,\eps_2,q) = \CF(a,\eps_1,\eps_2,q) = \eps_1 \eps_2 \CF^{\text{pert.}} + \eps_1 \eps_2 \log \left( \sum_{k \in \BZ}q^{k} \int_{\mathcal{M}_k} {\boldsymbol 1}\right) 
 \]
with $\lim_{\epsilon_{1,2}\to 0} \CF(a,\eps_1,\eps_2,q) = \CF_0(a)$ and ${\boldsymbol 1}$ denoting the fundamental class in $\CM_k$. The first higher order terms correspond presicely to graviational corrections which can be obtained by first expanding the free energy in terms of the equivariant parameters
 \be
\CF = \CF_0 + (\eps_1+\eps_2)H + (\eps_1 + \eps_2)^2 G + \eps_1 \eps_2 \CF_1 + \ldots
\ee 
and then by identifying \cite{Nakajima:2003uh}, 
 \begin{eqnarray}
 G &=& \frac{B}{3} \\
 \CF_1 &=& A - \frac{2B}{3}
 \end{eqnarray}
 Therefore, we need to expand the logarithm of the deformed partition function of the theory and obtain the coefficients $A$ and $B$. Then, by noting that
 \begin{eqnarray}
 A &=& \alpha \left( \tau_0, m \right) \left( \frac{du}{da} \right)^{\frac{1}{2}} \\[10pt]
 B &=& \beta\left( \tau_0, m \right) \Delta^{\frac{1}{8}}
 \end{eqnarray}
 we can obtain the functions $\alpha $ and $\beta$ by studying the theory at the neighborhoods of the cusps of $\CB$ \cite{Moore:1997pc,Shapere:2008zf}.

\subsection{The Grassmann and  photon path integral}
As we mentioned above, using the descent formalism and the UV to IR map we can obtain the observables of the low energy theory which lie in the $G$-equivariant $\overline{\CQ}$-cohomology. For us the two operators of interest correspond to operators supported on a point $p \in H_0(\bar{X})$ and on a two-cycle $\bfx \in H_2(\bar{X})$ since we work with four-manifolds that do not contain one- and three-cycles. The operators are denoted as $\mathsf{I}_0(p)$ and $\mathsf{I}_2(\bfx)$ respectively (see \cite{Moore:1997pc} for the precise definition of those operators). We want to understand the IR flow of the two UV operators. As it is known from \cite{Moore:1997pc} the UV to IR map is given by
\begin{eqnarray}
\mathsf{I}_0(p) &\to & 2pu \\[0.5em]
\mathsf{I}_2(\bfx) &\to & \tilde{\mathsf{I}}_{-}(\bfx) \
\end{eqnarray}
where for the surface operator supported on the arbitrary element $\bfx $ the IR operator is written as \cite{Tan:2009qq} (see \cite{Moore:1997pc} for the usual Donaldson-Witten theory).  
\be
\tilde{\mathsf{I}}_{-}(\bfx) = \frac{i}{\sqrt{2}\pi} \int_{\bfx} \left( \frac{1}{32}\frac{d^2u}{da^2}\psi \wedge \psi - \frac{\sqrt{2}}{4}\frac{du}{da}(\mathscr{F}_{-}+D) \right) + \frac{i}{2}\int_{\bfS}\alpha \frac{du}{da}\delta_{\bfy}^{-}.
\ee
Note that both $2pu$ and $\mathsf{I}_2(\bfx)$ are $\overline{\CQ}$-invariant sincd they arise from the canonical descent proceedure \cite{Tan:2009qq}. The last term in the previous equation is the additional term that this operator obtains in the presence of the surface defect and $\delta_{\bfy}^{-}$ corresponds to the ASD two-form that is Poincar\'e dual to $\bfy$. Tha latter is an arbitrary element in the middle homology of $X$. We will see that eventually the $u$-plane integral has no dependence on $\bfy \in H_2(X)$. In the limit $\alpha \to 0$ we return to the theory without the surface operator, that is equation (3.7) of \cite{Moore:1997pc}. Just like the theory without the surface operators develops a contact term operator dependence in the IR, due to the map 
\be
\mathsf{I}_2(\bfx)  \mathsf{I}_2(\bfy) \to  \tilde{\mathsf{I}}_{-}(\bfx) \tilde{\mathsf{I}}_{-}(\bfy) + \text{ contact term},
\ee
 the same will happen when we add the extra surface operators in the original UV theory. Therefore there is an analogue of the $\bfx^2 G(u)$ contact term, that is a term of the form ${\bfS}^2 H(u)$ that also has to be taken into account in the UV to IR map. In principle we should also include a term proportional to the intersection of $\bfx$ and $\bfS$. But, such a term would vanish since the ramified Donaldson invariants are defined for $\bfx \in H_2(\bar{X}) =H_2(X\backslash \bfS )$ and $\bfx$ is homologous to $\bfy$ and as a result $\bfy\cap \bfS =\bfx\cap \bfS  =0$. Taking all into account, as well as the fact we will be evaluating the theory on simply connected manifolds, $\pi_1(X)=0$, we therefore conclude that the UV to IR map is \cite[equation 5.8]{Tan:2009qq}
\be
\bra e^{p\mathsf{I}_0 + \mathsf{I}_2(\bfx)} \ket \to \bra e^{ 2pu - \frac{i}{4\pi} \int_{\bfx} \frac{du}{da}(\mathscr{F}_{-}+D) + \bfx^2G(u) + \tilde{\bfS}^2H(u)} \ket,
\ee
where $\tilde{\bfS} =  \frac{\pi i \alpha}{2}\bfS$. We already recognize two contact terms, unlike in \cite{Moore:1997pc, LoNeSha, Marino:1998bm, Korpas:2017qdo} where there is only one, with the second one here being precisely due to the presence of the surface defect. The contact terms are given by
\begin{eqnarray}
G(u) &=& \frac{1}{24}\left( 8u - E_2(\tau) \left( \frac{du}{da} \right)^2 \right), \\[0.5em]
H(u) &=& u ~p_2(u),
\end{eqnarray}
where the polynomial $p_2(u) = \sum_{n \in2 \BZ_{\geq 1}}a_n u^{-n} \in \BQ[u^{-1}]$ is chosen such that  it vanishes at the classical limit, $\lim_{u \to \infty} p_2(u) = 0$, and $p_2(-u)= p_2(u)$. In \cite{Tan:2009qq} the author chooses to use the simplest term, $u^{-2}$ with $a_2=\frac{1}{4}$. We can leave this polynomial arbitrary for the purposes of this paper. Note that none of the contact terms $G(u)$ and $H(u)$ have any singularities at the SW points.

To this end we will add a $\overline{\CQ}$-exact operator $\tilde{\mathsf{I}}_{+}(\bfx)$ to the partition function that is defined accrording to the lines of \cite{Korpas:2017qdo}  as
\begin{equation}
\tilde{\mathsf{I}}_{+}(\bfx) = -\frac{1}{4\pi} \int_{\bfx} \left\{ \overline{\CQ}, \frac{d\bar u}{d\bar a}\chi  \right\}.
\end{equation}
Since the physical operators of the (ramified or not) Donaldson-Witten theory belong to the $G$-equivariant cohomology of $X$ we are definitely allowed to do so as long as any insertion is $\overline{\CQ}$-exact. This operator first appeared in reference \cite{Moore:1998et} in the context of interpreting Witten type indices of bounds states in string theory in terms of CohFT integrals in various dimensions.

This $\overline{\CQ}$-exact operator needs to be treated very carefuly and studied to some detail. At this stage it is not clear wether it provides a well-defined observable of the low energy IR theory or not. So for this paper we will assume that indeed this is the case. Nevertheless, using number theoretic techniques, it can be shown that our assumption is valid and it is a well-defined operator. Paper \cite{Korpas:2018} studies $\tilde{\mathsf{I}}_{+}(\bfx)$ in detail and we refer the reader to it for further details.   

In the presence of the surface operator $\bfS$, it would seem natural though to add a second term, the one corresponding to the anti-holomorphic part of the last term in (5.11).  In order such an inclusion to not destroy the topological nature of the theory it has to be $\overline{\CQ}$-exact. Despite that we see that 
\bes
\left\{ \overline{\CQ}, \frac{d\bar u}{d \bar a} \delta_{\bfS} ^{+} \right\} =\sqrt{2}i \frac{d^2\bar{u}}{d\bar{a}^2} \eta \delta_{\bfS}^{+},
\ees
and we observe this operator is not $\overline{\CQ}$-exact (in other words we cannot write $\frac{i}{2}\int_{\bfS} \alpha \frac{d \bar{u}}{d \bar{a}} \delta_{\bfS}^+= \int_{\bfS} \{\overline{\CQ} , x \}$ for some combination of fields $x$). Therefore we will not add this term to the path integral. In principle we could cook up other $\overline{\CQ}$-exact operators such that to couple $\delta_{\bfS}^{+}$ to the theory but this is not something we will do in this paper. As a result our $\overline{\CQ}$-exact ``deformation" operator is 
\be
\tilde{\mathsf{I}}_{+}(\bfx,\bfS) = -\frac{i}{\sqrt{2}\pi} \int_{\bfx} \left( \frac{1}{2}\frac{d^2\bar{u}}{d\bar{a}^2}\eta \chi +\frac{\sqrt{2}}{4}\frac{d\bar{u}}{d\bar{a}}(\mathscr{F}_{+}-D) \right).
\ee 
The analysis is identical to the one of the unramified $u$-plane integral, the only differences being 
\begin{enumerate}
\item that we have to consider the (extended) vector bundle $\mathscr{E}$ instead of $\CE$ by making a choice of a lift of $\alpha$ from $\mathbb{T}$ to $\mathsf{t}$ (this point will become clear when we study the photon path integral), and 
\item the presence of the additional contact term $H(u)$.
\end{enumerate}

We are in position now to write down the precise form of the path integral together with the insertions that will compute the ramified Donaldson invariants
\begin{eqnarray}\label{PathInt}
Z_u(p,\bfx) = \int [\mathscr{D}\varphi] e^{ \SFO[\varphi]},
\end{eqnarray}
with 
\begin{eqnarray} \nonumber
\SFO[\varphi] & =& \SFO( a,\bar{a},A,\eta,\chi,D )  \\[0.2cm]  \nonumber
                & :=&\frac{i}{16\pi}\int_X  (\bar{\tau}|\mathscr{F}^{+}|^2+\tau|\mathscr{F}^{-}|^{2}) + \frac{\tau_2}{8\pi} \int_X da \wedge d\bar{a} - \frac{\tau_2}{8\pi} \int_X D\wedge *D \\[0.2cm]  \nonumber
                &+& \frac{\sqrt{2}i}{16\pi} \frac{d\bar{\tau}}{d\bar{a}}\int_X  \eta \chi \wedge (\mathscr{F}^{+}+D)  +\frac{i}{4}\eta_{\mathrm{eff.}}\int_X  \delta_{\bfS} \wedge \mathscr{F} + \bfx^2G(u) \\[0.2em]
                &+&  2pu + \bfx^2G(u) + \tilde{\bfS}^2H(u) \\[0.2cm]  \nonumber
                &+& \tilde{\mathsf{I}}_{-}(\bfx) + \tilde{\mathsf{I}}_{+}(\bfx).
\end{eqnarray}
The procedure to follow then is to try to perform the integration (\ref{PathInt}) directly. To this end we can eliminate the auxiliary self-dual two-form $D$ using its equation of motion
\[
D = -\frac{2 }{\tau_2} \mathrm{Im} \left( \frac{du}{da} \right)\bfx_{+} + \frac{\sqrt{2}i}{4\tau_2}\frac{d\bar{\tau}}{d\bar{a}} \eta \chi.
\]
The insertion of the surface defect does not alter the integration over the fermionic zero modes $\eta_0$ and $\chi_0$ of the theory and after performing the integration over them we obtain the same term as in \cite{Korpas:2017qdo} with the difference that instead of $F$ we have $\mathscr{F}$ in the argument of the quadratic form. Therefore modulo other terms that depend on $\mathscr{F}$ and will enter in the discussion of the photon path integral, we obtain
\[
\int [\mathscr{D}\eta_0\mathscr{D}\chi_0]e^{\mathrm{Grassman}} = \frac{\sqrt{\tau_2}}{4\pi} \frac{d\bar{u}}{d\bar{a}} \partial_{\bar{\tau}} (\sqrt{2\tau_2}B(\mathscr{F}-4\pi \bfb, \underline{J}) ).
\]
The reason we integrate over the Grassman zero modes only is something standard in CohFT and explained in detail in \cite[Section 2.3]{Moore:1997pc} but also in \cite{Laba05,  Korpas:2017qdo}. In the expression above $J \in H^2(X,\BR)$ and if $X$ is a K\"ahler surface we can view it as ${J}\in \CO_X(1)$. By $\underline{J}$ we denote the polarization $J$ normalized by $Q(J)$. Note that we have also defined the class
\[
\bfb = \frac{\textrm{Im}(\bfrho)}{\tau_2},
\]
for an elliptic variable $\bfrho \in H_2(X,\BC)\otimes \BM_{(1,0)}(\Gamma^0(4))$ and this class will appear in the photon path integral as well. Explicitly, we define it as
\[
\bfrho = \frac{\bfx}{2\pi}\frac{du}{da},
\]
where $\BM_{(a,b)}(\Gamma)$ denotes the space of modular forms of weight $(a,b)$ under $\Gamma$. This brings us to the photon path integral. Since the low energy effective theory corresponds geometrically to a connection on a U(1) line bundle $\CL$, the photon path integral will be a sum over all topological classes of such line bundles \cite{Witten:1995gf, Moore:1997pc}. This means that we need to sum over all the connected components of the Picard group $\mathrm{Pic}(X)$, in other words over all degrees\footnote{Over a K\"ahler surface $X$ each disconnected component of $\textrm{Pic}(X)$ is isomorphic to $H^1(X,\CO_X)/H^1(X,\BZ)$ which is a complex torus of dimension $b_1$. For a simply connected K\"ahler surface though $b_1=0$ and as a result each disconnected component is a point and the summation of the photon path integral is over $ \Lambda \cong \mathrm{Pic}(X) $. Note though that if $b_1>0$ there is an extra integration over each disconnected component of ${\rm Pic}(X)$ \cite{Marino:1998rg}, i.e., $\int_{{\rm Pic}(X)}d\psi \ldots$.}
Effectively we want to count possible first Chern classes $c_1(\CL)$ which physically correspond to magnetic fluxes. Therefore the path integral
\[
Z_A = \int [\mathscr{D}A] e^{-\int_X  \frac{i}{16\pi} (\bar{\tau}|\mathscr{F}^{+}|^2+\tau|\mathscr{F}^{-}|^{2}) }
\]
is simply a theta function that we will describe below. To this end we introduce the conjugacy class $\bfmu \in H_2(X,\BZ_2)$ such that $w_2(\mathscr{E}) = 2\bfmu + H_2(X,2\BZ)$ and $\frac{1}{4\pi}[\mathscr{F}] \in H^2(X,\BZ)+\bfmu$. Then the photon path integral is the theta function
\[
Z_{A} = \sum_{\tilde{\bfk} \in \Lambda+ \frac{1}{2} w_2(\mathscr{E})} e^{-\pi i \bar{\tau} \tilde{\bfk}_{+}^2 - \pi i \tau \tilde{\bfk}_{-}^2 }
\]
where we identify $\Lambda$ with the middle cohomology lattice $H^2(X,\BZ)$ as we discussed earlier while we allow the possibility to work with an $\mathrm{SO(3)}$ bundle which precisely means a shift to the sum by $\frac{1}{2} w_2(\mathscr{E})$. In the theta function we have defined  $[\mathscr{F}]/4\pi  = \bfk - \frac{\alpha}{2}\delta_{\bfS} := \tilde{\bfk}$. Of course this is not quite the correct form of the theta function since we need to include the standard $(-1)^{B(K_X,w_2(\mathscr{E}))}$ factor\footnote{Actually we can instead consider instead of $w_2(\mathscr{E})$ the canonical class $c_1(\CK_X) = K_X$.} which is obtained from  integrating out the massive fermionic degrees of freedom in the Coulomb branch $\CB$ \cite{Witten:1995gf}. Then, combining the theta function with the remainders of the Grassman integration we obtain the Siegel-Narain theta function
 \be
\label{Psi2}
\begin{split}
\tilde{\Psi}^J_{\bfmu}(\tau, {\bfrho}; \alpha
)&=e^{-2\pi\tau_2 \bfb_+^2}\sum_{\tilde{\bfk}
  \in \Lambda +\bfmu} \partial_{\bar
  \tau}\left(\sqrt{2\tau_2}B(\tilde{\bfk} + \bfb, \underline J)\right) (-1)^{B(\tilde{\bfk}, K_X)} \\ 
                                         &\times \,\exp\!\left( -\pi i \bar \tau \tilde{\bfk}_{+}^2 - \pi i \tau \tilde{\bfk}_{-}^{2}  -2\pi i  B(\tilde{\bfk}_{+} , \bar{{ \bfrho}}) - 2\pi i B(\tilde{\bfk}_{-},{\bfrho})\right)  \\                                        
                                         &\times \, \exp\!\left(-2\pi i B(\tilde{\bfk}, \frac{\eta}{2}\delta_{\bfS}) \right),
\end{split}
\ee  
In general, unless we want to stress the dependence of the $\mathtt{t}$-lift (choice of $\alpha$), we will omit it from the functions it appears. It is straight forward to see the relation of the Siegel-Narain function for the ramified theory to the unramified one, that is
\be
\lim_{(\alpha,\eta)\to (0,0)} \tilde{\Psi}^J_{\bfmu}(\tau, \bfrho) = \Psi^J_{\bfmu}(\tau, \bfrho),
\ee
where the latter is the Siegel-Narain theta function of the theory without the presence of the surface defect $\bfS$ \cite{Korpas:2017qdo} 
 \be
\label{Psi0}
\begin{split}
\Psi^J_{\bfmu}(\tau,{\bfrho}
)&=e^{-2\pi\tau_2 \bfb_+^2}\sum_{{\bfk}
  \in \Lambda +\bfmu} \partial_{\bar
  \tau}\left(\sqrt{2\tau_2}B({\bfk} + \bfb, \underline J)\right) (-1)^{B({\bfk}, K_X)} \\ 
                                         &\times \,\exp\!\left( -\pi i \bar \tau {\bfk}_{+}^2 - \pi i \tau {\bfk}_{-}^{2}  -2\pi i  B({\bfk}_{+} , \bar{{ \bfrho}}) - 2\pi i B({\bfk}_{-},{\bfrho})\right).  \
\end{split}
\ee  
Note that the $\eta$ that appears in (\ref{Psi2}) is the ``magnetic charge" associated to $\alpha$ and not the Grassman valued scalar field of course. 

Taking a closer look at $\tilde{\Psi}^J_{\bfmu}(\tau, {\bfrho})$ and requiring that it has the correct modular behaviour (the discussion of which we postpone for subsection \ref{modularity} ) in order the integrand of the ramified $u$-plane integral to be modular invariant, forces  $\alpha \in \BZ$ for $\mathrm{SO}(3)$ bundles and $\alpha \in 2\BZ$ for $\mathrm{SU}(2)$ bundles. The last term in (\ref{Psi2}) is equal to one therefore and as a result of  requiring that (\ref{Psi0}) has the correct modular properties, we pick specific allowed surface operators, that is pairs $(\alpha,\eta)$, shown in Figure \ref{charges} below. 

\begin{figure}[!htb]
    \centering
    \begin{minipage}{.5\textwidth}
        \centering
        \includegraphics[scale=0.65]{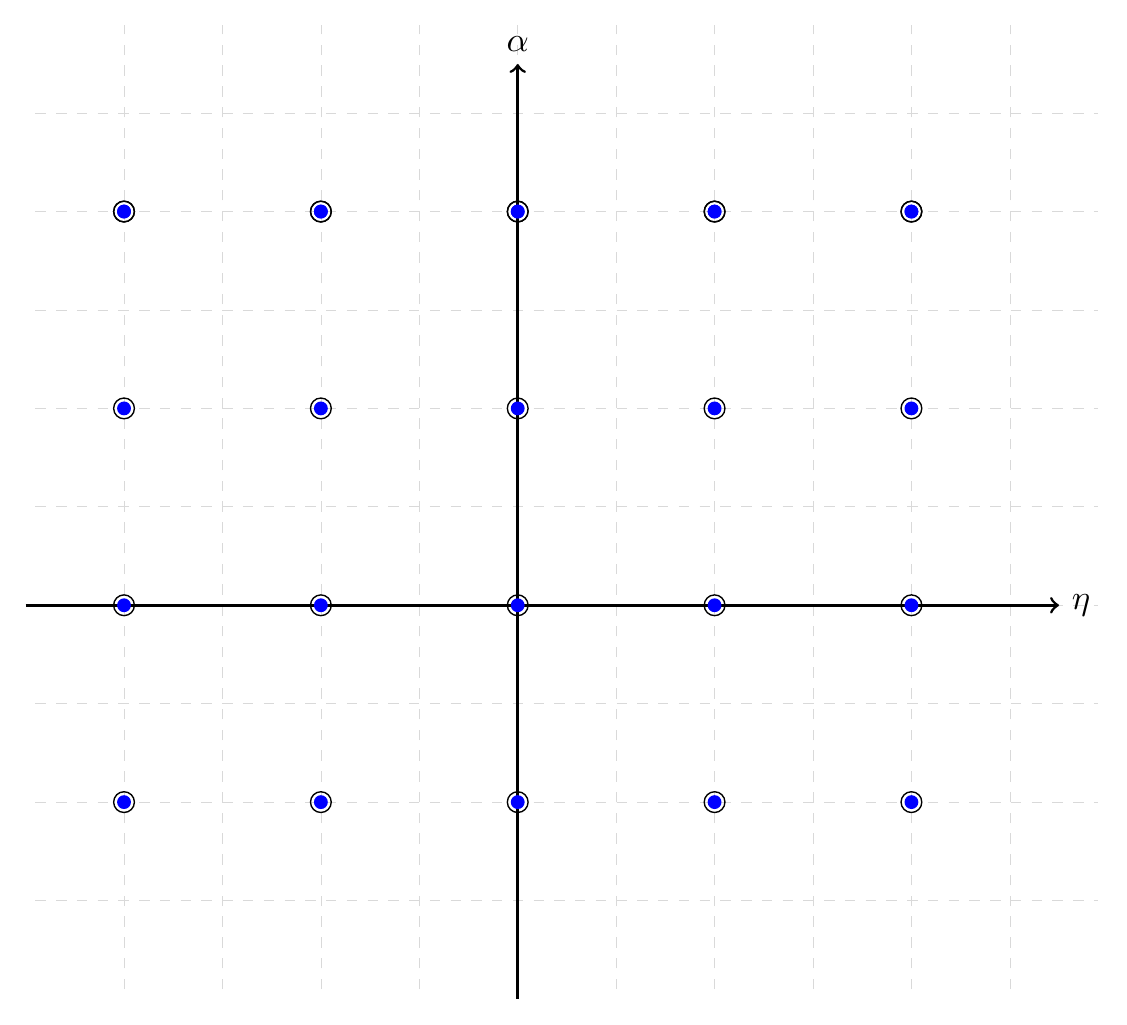}
    \end{minipage}%
    \begin{minipage}{0.5\textwidth}
        \centering
        \includegraphics[scale=0.65]{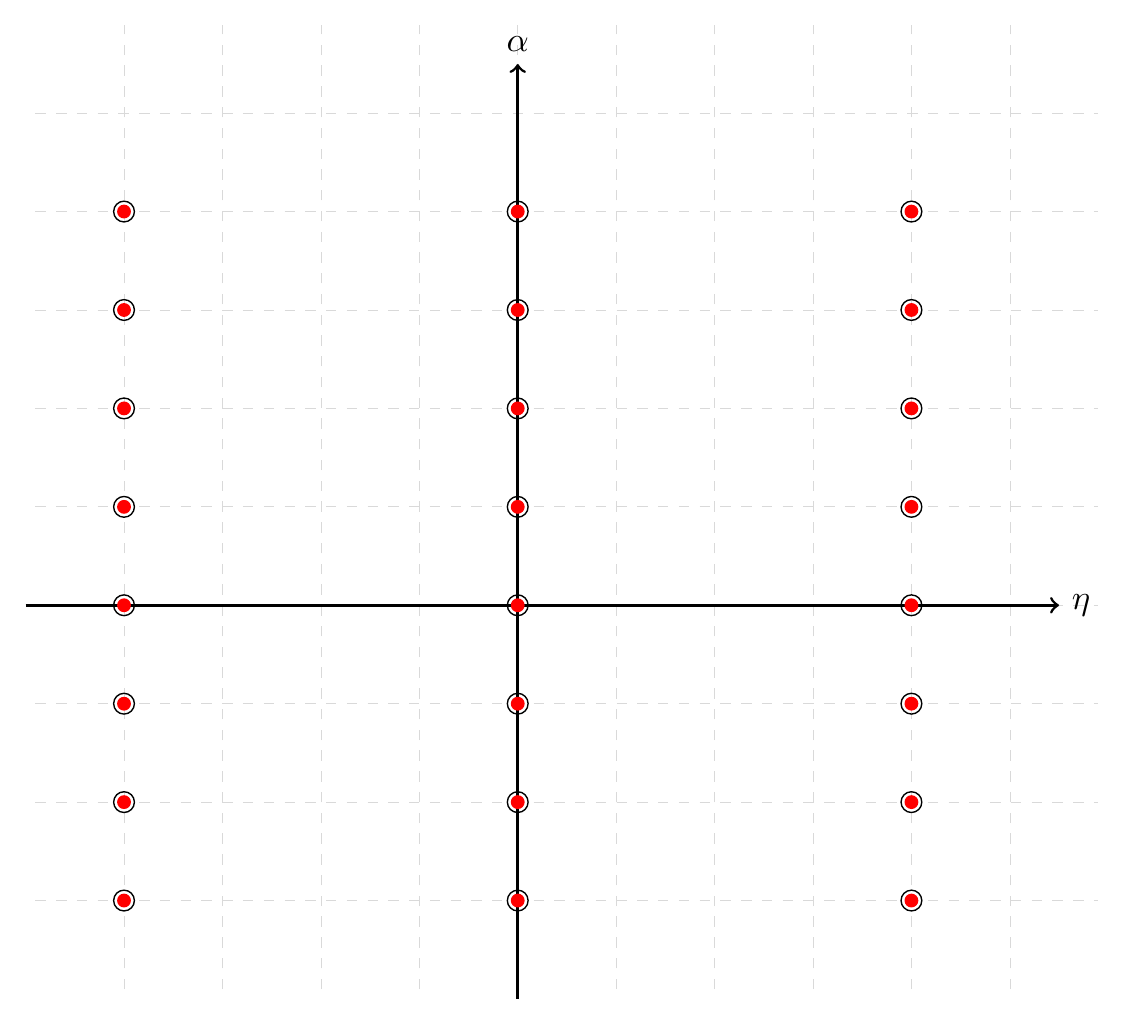}
    \end{minipage}
       \label{SOps}
    \caption{Electic and magnetic charges  of the surface operator for gauge group ${\rm SU}(2)$ in blue and ${\rm SO(3)}$ in red. Both lattices are integer. \label{charges}}
\end{figure}

\subsection{The ramified $u$-plane integral}
We are in position now to write down the $u$-plane integral $Z_u (p,\bfx):= \tilde{\Phi}_{\bfmu}^{J}(p,\bfx)$ for the theory in the presence of surface defects (the change of notation is to emphasize the dependence on both $\bfmu$ and $J$). Taking into account the measure factor $\nu(\tau)$, the point and contact term operators associated to $p, \bfx, \bfS$ and the Siegel-Narain theta function we have
\be \label{Phi}
\begin{split}
\tilde{\Phi}_{\bfmu}^{J}(p,\bfx) &= \int_{\CB} da \wedge d\bar{a} \, \nu(\tau) e^{2pu + \bfx^2 G(u) + \tilde{\bfS}^2 H(u)} \tilde{\Psi}_{\bfmu}^J(\tau, \bfrho).
\end{split}
\ee
The integration domain of the integral above is the Coulomb branch $\CB$ which is identified with $\BH/\Gamma^{0}(4) \cong \BC \BP^1 \backslash \{\pm 1, \infty \}$. It is more natural to make a coordinate transformation for the measure therefore and write it in terms of the complexified gauge coupling. By defining 
\[
\tilde{\nu}(\tau) := \frac{da}{d\tau} \nu(\tau),
\]
we can write (\ref{Phi}) as the following modular integral

\be \label{Phi2}
\begin{split}
\tilde{\Phi}_{\bfmu}^{J}(p,\bfx) &= \int_{\BH/\Gamma^{0}(4)} d\tau \wedge d\bar{\tau} \, \tilde{\nu}(\tau) e^{2pu + \bfx^2 G(u) + \tilde{\bfS}^2 H(u)} \tilde{\Psi}_{\bfmu}^J(\tau, \bfrho).
\end{split}
\ee
Let us make a remark at this point. The Siegel-Narain theta function is not a holomorphic function, i.e. it depends both on $\tau$ and $\bar{\tau}$, as clearly seen in (\ref{Psi0}), and this is a very crucial point that will allow the localization of the integral (\ref{Phi2}) to specific points of the Coulomb branch.

\subsection{Modularity of the ramified $u$-plane integrand} \label{modularity}
In order the ramified $u$-plane integral (\ref{Phi2}) to make sense it has to be modular invariant under $\Gamma^0(4)$. In (\ref{Phi2}) the measure $d\tau \wedge d\bar{\tau}$ has modular weight $(-2,-2)$ under $\Gamma^0(4)$. As a result we have to require that the integrand transforms as a modular form of weight $(2,2)$ in order to obtain a single-valued quantity. Earlier, we characterized $\bfrho$ as an elliptic variable and this is due to the fact that under the congruent subgroup at hand $\frac{du}{da}$ is a modular form of weight one. In specific, under an $S$ transformation $ \tau \mapsto \frac{\tau}{\tau+1}$ we have
\[
\frac{du}{da} \mapsto \frac{1}{\tau+1}\frac{du}{da}.
\]
Using the standard properties of Jacobi theta functions we showed in \cite{Korpas:2017qdo} that $\bfrho$ transforms under the two generators of $\Gamma^0(4)$ as
\be
\begin{split}
\bfrho(\tau+4) &= -\bfrho(\tau) \\[0.3cm]
\bfrho \left(\frac{\tau}{\tau+1}\right) & = \frac{\bfrho(\tau)}{\tau+1}
\end{split}
\ee
Using these transformations and also the fact that $\bfk = \bfl + \bfmu + \frac{K_X}{2}$ (where we can perform the shift by $\frac{K_X}{2}$ since $K_X$ is a characteristic vector of the lattice $\Lambda$ as follows from the Hirzebruch-Riemann-Roch theorem\footnote{A characteristic vector $K \in \Lambda$ is defined as following: for $\bfv \in \Lambda$ we have $\bfv^2 = B(\bfv, K) \mathrm{mod}2$. It is a fact that a characteristic vector always exists. The Hirzebruch-Riemann-Roch theorem, or Riemann-Roch therem for surfaces, states that for a complex surface $X$ and for a line bundle $L=\CO(D)$, where $D$ is an effective divisor, we have \[ \chi(\CO_X(D)) = \frac{1}{2} B(D, D-K_X) + \chi(\CO_X). \] Since the Euler characteristic of any line bundle is an integer, by taking their difference and multiplying by two we get that $B(D, D-K_X) \in 2\BZ$. }) we can find the $T$ and $S$ transformations of $\Psi_{\bfmu}^{J}(\tau,\bfrho)$. For details see \cite{Korpas:2017qdo}. 

\subsubsection*{The $T$ transformation}
In order to find the $T$ transformation $\TT \mapsto \TT+4$ for $\Gamma^0(4)$ we will apply a $\TT \mapsto \TT+1$ transformation four times. We will also allow a generic shift $\bfmu \mapsto \bfmu + \frac{K_X}{2}$. Therefore we have
\begin{eqnarray}
\tilde \Psi_{\bfmu + \frac{K_X}{2}}^J(\TT+1,\bfrho) &=& e^{\pi i ( \bfmu^2 - \frac{K_X^2}{4} )} \tilde \Psi_{\bfmu + \frac{K_X}{2}}^J(\TT,\bfrho + \bfmu),
\end{eqnarray}
and repeating four times yields
\begin{eqnarray}
\tilde \Psi_{\bfmu + \frac{K_X}{2}}^J(\TT+4,\bfrho) &=& e^{4\pi i ( \bfmu^2 - \frac{K_X^2}{4} )} \tilde \Psi_{\bfmu + \frac{K_X}{2}}^J(\TT,\bfrho + 4\bfmu).
\end{eqnarray}
We can get rid off the shift $\frac{K_X}{2}$ and also by noting that $B(\bfk, 4\bfmu) \in 2\BZ$ we finally obtain 
\begin{eqnarray} \label{5.27}
\tilde \Psi_{\bfmu }^J(\TT+4,\bfrho) &=& e^{\pi i B(4\bfmu, K_X)} \tilde \Psi_{\bfmu}^J(\TT,\bfrho ) .
\end{eqnarray}
Note here that we have not treated the transformation of $\bfrho(\tau)$ yet. Since $K_X$ is a characeristic vector, we have that $\bfl^2 + B(l,K_X) \in 2\BZ$ for any vector $\bfl \in \Lambda$. Therefore, the exponential we see in (\ref{5.27}) can be written as 
\[
(-1)^{Q(2\bfmu) + B(\bfmu,K_X) - 3B(2\bfmu, K_X)} = e^{-6\pi i B(\bfmu, K_X)}
\]
since $Q(2\bfmu) + B(\bfmu,K_X) \in 2\BZ$. As a result we can write (\ref{5.27}) taking into account the transformation of $\bfrho$ under $\Gamma^0(4)$ we obtain
\begin{eqnarray} \label{Ttransf}
\tilde \Psi_{\bfmu }^J(\TT+4,-\bfrho) &=& -e^{2\pi i B(\tilde{\bfk}, K_X)} \tilde \Psi_{\bfmu}^J(\TT,\bfrho ) .
\end{eqnarray}

\subsubsection*{The $S$ transformation}
Similarly, for the $S$ transformation we can first perform a $\tau \mapsto -\frac{1}{\TT}$ transformation and then generalize. We have therefore
\be
\begin{split}
\tilde \Psi_{\bfmu + \frac{K_X}{2}}^J \left( -\frac{1}{\TT}, \frac{\bfrho}{\TT} \right) &= -i(-i\TT)^{\frac{n}{2}}(i\bar{\TT})^2 e^{-\frac{\pi i \bfrho^2}{\TT}+\pi i \frac{K_X^2}{2}}(-1)^{B(\bfmu,K_X)} \tilde{\Psi}_{\frac{K_X}{2}}^J(\TT, \bfrho - \bfmu),
\end{split}
\ee
and similarly as for the $T$ transformation, by repeating four times this procedure we obtain
\be
\tilde \Psi_{\bfmu + \frac{K_X}{2}}^J \left( \frac{\TT}{\TT+1}, \frac{\bfrho}{\tau+1} \right) =  (\bar{\TT}+1)^2(\TT+1)^{\frac{b_2}{2}} e^{-\frac{\pi i \bfrho^2}{\TT+1} + \frac{\pi i \sigma(X)}{4}} \tilde{\Psi}_{\bfmu}^{J}(\tau, \bfrho),  
\ee
where we use the fact that for simply connected four-manifolds we have $K_X^2 = \sigma(X) + 8$.

\subsubsection*{The rest of the terms}
As for the rest of the terms, that is the contact terms as well as the measure factor the analysis is identical to \cite{Korpas:2017qdo} without any modifications. The contact term transforms as 
\be
\begin{split}
e^{\bfx^2 G(\tau+4)} &= e^{\bfx^2G(\tau)},\\
e^{\bfx^2 G(\frac{\tau}{\tau+1})} &= e^{\bfx^2G(\tau) + \frac{\pi i }{\tau+1}\bfrho^2}.
\end{split}
\ee
Note that the function $H(u)$ is modular invariant and thus its transformations are trivial. As a result we see that indeed the integrand has the desired (2,2) weight under $\Gamma^0(4)$. This is in perfect agreement with the proof of Tan that the ramified $u$-plane integrand without the $\overline{\CQ}$ insertion is indeed modular invariant. The fact that our integrand is modular invariant is no surprise since the insertion of the supersymmetric surface operator supported on $\bfS$ does not contribute to the modularity properties of the integrand. In Table \ref{Table} we summarize the modular weights of the various functions that appear in our considerations.

\subsection{The ramified $u$-plane integral as a total derivative} \label{ThetaSection}
In \cite[Section 4]{Korpas:2017qdo} we expressed the integrand of the $u$-plane integral $\Phi_{\bfmu}^J(p,\bfx)$ of the Donaldson-Witten theory without surface operators in terms of the total $\bar{\TT}$ derivative of a non-holomorphic function $\widehat{\CH}$. This function is the modular completion of a mock modular form $\CH(\tau)$ whose shadow is the Siegel-Narain theta function $\Psi_{\bfmu}^J$. As we will see, these ideas can be applied for the ramified $u$-plane integral as well in a straight forward manner . 

\begin{figure}[h] \centering
\includegraphics[scale=1.3]{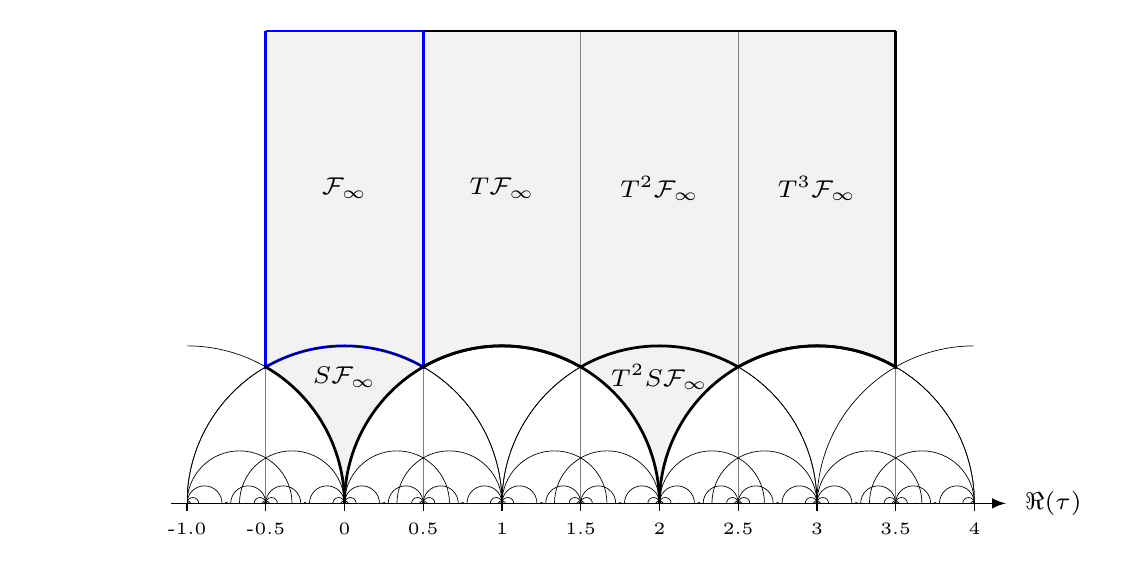}
\caption{\small The fundamental domain of ${\rm SL}(2,\BZ)/\Gamma^0(4)$. We denote by $\CF_{\infty}$ the fundamental domain of ${\rm SL}(2,\BZ)$. \label{Fig4} }
\end{figure}

Let us begin by discussing the domain of integration $\BH/\Gamma^0(4)$ which is the union of six images of the fundamental domain of $\mathrm{SL}(2,\BZ)$ \cite{Moore:1997pc,Tan:2009qq,Korpas:2017qdo} This is depicted in Figure \ref{Fig4}. To be more precise $\BH/\Gamma^0(4)$ can be written as following
\bes
\BH/\Gamma^0(4) \cong  \left( \CF_{\infty} \cup T \CF_{\infty} \cup T^2 \CF_{\infty} \cup T^3\CF_{\infty}\right) \cup (S \CF_{\infty} \cup T^2S\CF_{\infty}).
\ees
The first four domains correspond to the cusp at infinity $i\infty$ of the semiclassical approximation. The next two domains correspond to the monopole point $\tau =0$ and dyon point $\tau=2$ which map to $-1$ and $+1$ in the Coulomb branch $\CB$. These domains are clearly seen in Figure \ref{Fig4} below.

 Integrals over $\BH/\Gamma^{0}(4)$ of modular invariant integrants of the form $d\tau \wedge d\bar{\tau} ~h(\tau, \bar{\tau})$ can be evaluated, in special cases in a quite straight forward way. These cases involve integrands that can be expressed as the total anti-holomorphic derivative to $\bar{\tau}$ of very specific function $\widehat{\CH}$ with the property 
\be \label{F}
 \frac{\partial \widehat{\CH}(\tau, \bar{\tau}) }{\partial \bar{\tau}} = h(\tau,\bar{\tau}).
\ee
The function $\widehat{\CH}(\tau, \bar{\tau}) $ is a modular form of $(2,0)$ and it is the modular completion of some (holomorphic) mock modular form of weight $(b_2/2,0)$. Then integrals such as the ramified $u$-plane integral (\ref{Phi2}) can be evaluated by relating them to integrals over $\CF_{\infty}$ and as we will see they localize to the cusps of $\BH/\Gamma^0(4)$ \cite[Appendix C]{Korpas:2017qdo}. To demonstrate this,  let us assume we want to evaluate the following integral,
\be \label{IntegralLocalizes}
I = \int_{\BH/\Gamma^0(4)} d\tau \wedge d\bar{\tau}~ h(\tau,\bar{\tau}).
\ee
What we can do is the following. Instead of trying to calculate $I$, we can compute the simpler integral $I_{\infty}$ defined as
\be \label{IntegralLocalizes}
I_{\infty} = \int_{\CF_{\infty}} d\tau \wedge d\bar{\tau}~ h(\tau,\bar{\tau}).
\ee
Using the property of $f(\tau, \bar{\tau})$ (\ref{F}) and Stoke's theorem, the integral (\ref{IntegralLocalizes}) can be written as a one-dimensional integral over $\tau$,
\be
I_{\infty} = \int^{i\infty} d\tau ~ \widehat{\CH}(\tau,\bar{\tau}).
\ee
where the integral in the rhs is evaluated at the cusp $i \infty$. The result of this integration is simply the $q^0$ coefficient of the $q$-series expansion of $\widehat{\CH}$ near the cusp $i\infty$,
\be
I_{\infty} = \Big[ \widehat{\CH}(\tau,\bar{\tau}) \Big]_{q^0}.
\ee
Then, it is quite straight forward to evaluate $I$ using inverse mapping, that is by repeating the $I_{\infty}$ integration for each of the other five $\CF_{\infty}$ domains of $\BH/\Gamma^0(4)$. The final result is
\be \label{RESULT}
I = 4\Big[\widehat{\CH}(\tau,\bar{\tau})\Big]_{q^{0}} + \Big[ S \CF_{\infty} \Big]_{q^{0}} +  \Big[ T^2S \CF_{\infty} \Big]_{q^{0}}  .
\ee
We can apply this method very nicely to the ramified $u$-plane integral (\ref{Phi2}). Let us first define the holomorphic function
\be
\tilde{f}(\tau) = \tilde{\nu}(\tau) e^{2pu + \bfx^2G(u) + \tilde{\bfS}^2 H(u) }.
\ee
Then the ramified $u$-plane integral can be written as 
\be \label{Phi}
\begin{split}
\tilde{\Phi}_{\bfmu}^{J}(p,\bfx) &= \int_{\BH/\Gamma^0(4)} d\tau \wedge d\bar{\tau} ~\tilde{f}(\tau) \tilde{\Psi}_{\bfmu}^J(\tau, \bfrho).
\end{split}
\ee
As we will show below, there exists a mock theta function $\varTheta(\tau)$ whose modular completion satisfies Equation (\ref{F}) and fits nicely in this context.

Let us briefly discuss how such a function appears in ordinary Donaldson-Witten theory. In \cite{Korpas:2017qdo}, (see also \cite{Korpas:2018}) for a detailed analysis) we were able to write the Siegel-Narain theta function $\Psi_{\bfmu}^{J}(\tau, \bfrho)$ of the $u$-plane integral as an anti-holomorphic derivative of an indefinite theta function $\widehat{\Theta}_{\bfmu}^{JJ'}(\tau, \bfrho)$ as 
\be \label{mainIdea}
\partial_{\bar{\tau}}\widehat{\Theta}_{\bfmu}^{JJ'}(\tau, \bfrho) = \Psi_{\bfmu}^J(\tau, \bfrho).
\ee 
This function, that is defined (see Equation \ref{RealTheta}) using a class $J' \in H^2(X,\BR)$ with the property $Q(J')=0$, reads
\be 
\label{hatTheta_sec}
\begin{split}  
\widehat \Theta^{JJ'}_{\bfmu}\!(\tau,\bfrho)&=\sum_{\bfk\in \Lambda+\mu} 
\tfrac{1}{2}\left\{ E(\sqrt{2\tau_2}\,B(\bfk+\bfb, \underline J))-\sgn(\sqrt{2\tau_2}\,B(\bfk+\bfb,J'))\right\} \\
& \times (-1)^{B(\bfk,K_X)} q^{-\frac{\bfk^2}{2}} e^{-2\pi i B(\bfk, \bfrho)}.
\end{split}  
\ee  
We see that the definition of $\widehat \Theta^{JJ'}_{\bfmu}$ involves the function $E(t):\mathbb{R}\to [-1,1]$ which is a reparametrization of the error function,
\begin{equation}
E(t) = 2\int_0^t e^{-\pi u^2}du = \text{Erf}(\sqrt{\pi}t),
\end{equation}
and $\underline J=J/ \sqrt{Q(J)}$ is the normalization of $J\in H^2(X,\BR)$, corresponding to a period point, as before. In Appendix \ref{AppA} we give the definition in more detail and discuss some standard properties of the indefinite theta functions.  Let us summarize all relevant weights under modular transformation for the functions of interest in the following table.
\begin{table}[h] 
 \renewcommand*{\arraystretch}{1.4}
\centering
\begin{tabular}{|| l@{\hspace{ 1in}}l||} \hline
Modular form & Mixed weight under $\Gamma^0(4)$   \\ \hline
$d \tau \wedge d\bar{\tau}$ & $(-2,-2)$  \\
 $\tilde{\nu}(\tau)$ & $ (2-b_2/2,0) $ \\
 $\sqrt{\tau_2}$ & $(-1/2, -1/2)$ \\ 
 $ \Psi_{\bfmu}^{J}(\tau,\bfrho) $ &  $(b_2/2,2)$   \\
 $\widehat{\Theta}_{\bfmu}^{JJ'}(\tau,\bfrho)$ & $(b_2/2,0)$\\ \hline
  $\partial_{\bar{\tau}}$ operator & raises $(\ell,0)$ to $(\ell,2)$ \\ \hline \hline
\end{tabular}
\caption{The mixed $\Gamma^0(4)$ weights of various modular forms that appear in the ramified $u$-plane integral. \label{Table}}
\end{table}
\noindent
Now we are ready to use the indefinite theta function (\ref{hatTheta_sec}) for our purposes. Let us define for convenience the function
\be
f(\tau) =  {\nu}(\tau) e^{2pu + \bfx^2G(u) },
\ee
that is the holomorphic function that multiplies the Siegel-Narain theta function $\Psi_{\bfmu}^{J}(\tau, \bfrho)$. Then we find that the function we seek in Equation (\ref{IntegralLocalizes}) is
\be
\widehat{\CH}_{\bfmu}(\tau,\bfrho) = f(\tau) \, \widehat{\Theta}_{\bfmu}^{JJ'}(\tau,\bfrho).
\ee
and following (\ref{RESULT}) the result of the $u$-plane integral can simply be written as
\be
\Phi_{\bfmu}^J(\tau,\bfrho) = 4 \Big[ f(\tau) \widehat{\Theta}_{\bfmu}^{JJ'}(\tau,\bfrho) \Big]_{q^0} +  \Big[ S \CF_{\infty} \Big]_{q^0} + \Big[ T^2S \CF_{\infty} \Big]_{q^0}.
\ee
For the theory with surface defects though, we already see that the Siegel-Narain theta function is defined with respect to the vector bundle $\mathscr{E} \to X$ (it also contained an extra term that depends on $\bfS$ but we showed this term to be equal to one due to the fact that $\alpha$ has to be an integer for $\tilde{\bfk}$ to belong to the lattice).  Using this fact we are now able to rewrite the integrand of (\ref{Phi2}) replacing $\tilde{\Psi}_{\bfmu}^J(\tau,\bfrho)$ with the anti-holomorphic derivative of the indefinite theta function 
\be 
\label{hatTheta_sec2}
\begin{split}   
\widehat \varTheta^{JJ'}_{\bfmu}\!(\tau,\bfrho)=\sum_{\bfk\in \Lambda+\mu} &
\tfrac{1}{2}\left( E(\sqrt{2\tau_2}\,B(\tilde{\bfk}+\bfb, \underline J))-\sgn(\sqrt{2\tau_2}\,B(\tilde{\bfk} +\bfb,J'))\right)\\
& \times (-1)^{B(\tilde{\bfk}, {K}_X)} q^{-\frac{\tilde{\bfk}^2}{2}} e^{-2\pi i B(\tilde{\bfk}, \bfrho)}.
\end{split}  
\ee  
Note that this function explicitly depends on the choice of lift of the connection to $\mathsf{t}$ via $\alpha$. 
As a result, the ramified $u$-plane integral (\ref{Phi2}) can be written as 
\be \label{contourInt}
\tilde{\Phi}_{\bfmu}^J(p,\bfx,\bfS) =\sum_{\partial (\CB)}  \oint du ~  \left( \frac{d\tau}{du} \right) \widehat{\CH}(\tau,\bfrho;\alpha),
\ee
where the function $\widehat{\CH}$ we were seeking reads for the ramified theory as following (note that we stress the dependence on $\alpha$),
\be
\widehat{\CH}(\tau,\bfrho;\alpha) =  \tilde{f}(\tau;\alpha) \, \widehat{\varTheta}_{\bfmu}^{JJ'}(\tau, \bfrho;\alpha).
\ee
\begin{figure}[h] \centering
\includegraphics[scale=0.95]{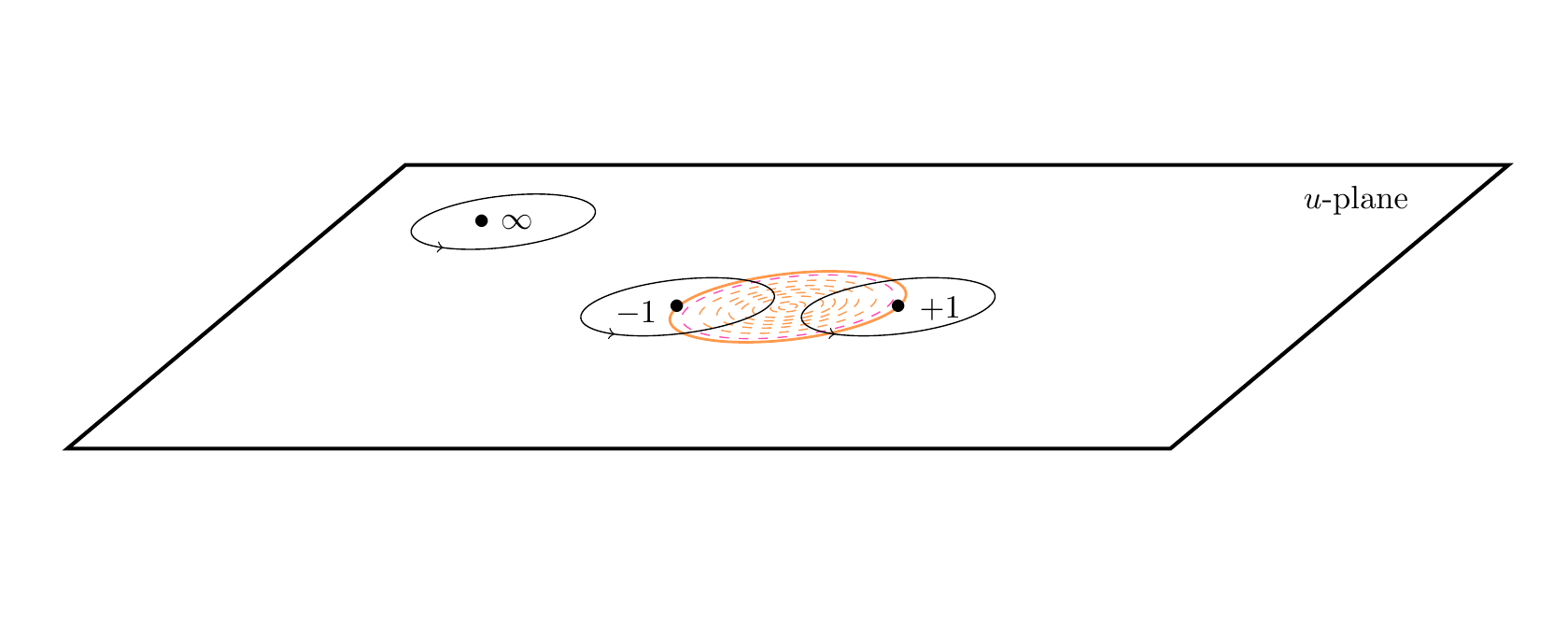}
\caption{\small Equation (\ref{contourInt}) shows that the contributions to the ramified $u$-plane integral come from the sum of thee contour integrals, from the boundaries of $\CB$, whose integrand contains the modular completed function $\widehat{\CH}$. In orange we depict the strong coupling region of the $u$-plane. The points $-1$ and $+1$ correspond to the monopole and dyon points respectively. The point at $\infty$ corresponds to the semi-classical region.}
\end{figure}
Therefore the ramified $u$-plane integral can be evaluated by the following formula (where we substitute the whole expression for $\tilde{f}$),
\be \label{Result}
\tilde{\Phi}_{\bfmu}^{J}(p,\bfx,\bfS) = 4\Big[ \tilde{\nu}(\tau) e^{2pu + \bfx^2G(u) + \tilde{\bfS}^2H(u)} \widehat{\varTheta}_{\bfmu}^{JJ'}(\tau,\bfrho;\alpha)  \Big]_{q^0}
 +  \Big[ S \CF_{\infty} \Big]_{q^0}  + \Big[T^2 S \CF_{\infty} \Big]_{q^0}.
 \ee
At the $\tau \to i\infty$ limit the error function becomes the sign function and as a result we can substitute the non-holomorphic indefinite theta function $\varTheta_{\bfmu}^{JJ'}(\tau,\bfrho)$ with the holomorphic indefinite theta function ${\varTheta}_{\bfmu}^{JJ'}(\tau,\bfrho)$. See Appendix \ref{AppA} and \cite[Appendix B]{Korpas:2017qdo} for details. Note that we have written these functions in italics in order to stress that they are (slightly) different that the functions defined in \cite{Korpas:2017qdo} due to the extended gauge bundle and the dependence on $\alpha$ and $\bfS$.  For four-manifolds which admit a metric of positive scalar curvature formula (\ref{Result}) reduces to just the first summand which in turn completely determines the ramified Donaldson invariants. For example, this is valid for specific examples of K\"ahler surfaces of Kodaira dimension $-\infty$ that (in addition) are simply connected. Such surfaces are include the Hirzebruch surfaces $\BF_l$, the projective plane $\BC \BP^2$ and some blow-ups of it. Actually, even for the computation of the usual Donaldson invariants for $\BC \BP^2$ for which $b_2=1$ and the class of the period point is proportional to the hyperplane class $H$, one needs to use the blow up $\widehat{\BC \BP^2}$ in order to apply the indefinite theta functions in the evaluation of the $u$-plane integral and of course this is also true for the ramified theory with the embedded surfaces. For four-manifolds that do not satisfy this criterion we should also take into account the contributions ${Z}_{\rm SW}$ from the Seiberg-Witten points of the Coulomb branch $\CB$ as we mentioned earlier and these contributions were derived in \cite{Tan:2009qq}.

\subsection{Wall-crossing formula} \label{WCformula}
It is well known that Donaldson invariants for four-manifolds with $b_2^+=1$ are only piecewise invariants. They depend on the chamber of $J$ under consideration. The same phenomenon is true for the ramified Donaldson invariants as well \cite{kronheimer19952}. This means that Donaldson invariants jump discontinuously as we move across \emph{walls} that divide the space of self-dual two-forms into various chambers. In each of those chambers Donaldson invariants are constant under smooth variations of the metric. The wall-crossing formula was derived in the context of the $u$-plane integral in \cite{Moore:1997pc}. Note that similar behaviour has recently been observed for the $u$-plane integral of the AD3 theory \cite{Moore:2017cmm}. The wall-crossing formula prescribes this discontinuous change of $\Phi_{\bfmu}^J$ under the variation of a metric with period point $J_0 \in H^2(X,\BR)$ to another metric with period point $J_1\in H^2(X,\BR)$. If these two period points belong to the same chamber the result vanishes of course. Let us recall that a wall is defined as following. Firsr we consider the ``forward" positive cone $V_{+}:= \{ J \in H^2(X,\BR)| Q(J)>0 \}$. Then, any $\xi \in H^2(X,\BZ)$ such that $Q(\xi)<0$ defines a wall in $V_{+}$ by
\be
W_{\xi} := \{ J \in V_{+} | B(\xi, J) = 0 \}.
\ee
The complement of the walls in the positive cone are the chambers. Due to the presence of the surface defect the walls are defined as following for the theory with the defects
\[
W_{\tilde{\bfk};\alpha} := \{ J \in V_{+}| B( \tilde{\bfk},J)=0 \},
\]
and when comparing to the unramified theory this tells us that the walls are shifted in $H^2(X,\BR)$ and this shifting explicitly depends on the choice of lift of the maximal torus $\mathbb{T}$ to $\mathsf{t}$ via $\alpha$.
Using the same argumentation as in \cite{Korpas:2017qdo}, but for the ramified theory, the difference of the Coulomb branch between two neighboring chambers is given by a term $\Delta \tilde{\Phi}_{\bfmu}^{J_1J_0} =\tilde{\Phi}_{\bfmu}^{J_1}-\tilde{\Phi}_{\bfmu}^{J_0} $  which reads
\be
\Delta \tilde{\Phi}_{\bfmu}^{J_1J_0} = \int_{\BH /\Gamma^{0}(4)} d\tau \wedge d\bar{\tau} ~ \tilde{f}\, (\tilde{\Psi}_{\bfmu}^{J_1}-\tilde{\Psi}_{\bfmu}^{J_0} ),
\ee
with the contribution from the cusp at $i\infty$ giving the following result
\be
\Delta \tilde{\Phi}_{\bfmu}^{J_1J_0}(p,\bfx) = 4 \Big[ \tilde{\nu}(\tau) e^{2pu + \bfx^2G(u) + \tilde{\bfS}^2H(u)}\widehat{\varTheta}_{\bfmu}^{J_1J_0}(\tau, \bfrho;\alpha) \Big]_{q^0}.
\ee
This can be seen as the difference of the ramified $u$-plane integral for two metrics corresponding to $J_0$ and $J_1$. Note that here both $J_0$ and $J_1$ are period points in $H^2(X,\BR)$ and as a result the indefinite theta function $\widehat{\varTheta}$ contains an error function for both $J_0$ and $J_1$ (see Equation \ref{BB2}). It is trivial to show that this formula reduces to formula (4.11) of \cite{Korpas:2017qdo} in the limit $(\alpha, \eta) \to 0$. Finally note that in \cite{Tan:2009qq} it is also shown that the wall-crossing formula of $Z_u= \Phi_{\bfmu}^J$ for the SW points $+1$ and $-1$ of the Coulomb branch $\CB$ cancel the contribution that can arise from the call-crossing of the ramified Seiberg-Witten invariants $Z_{\rm{SW}}$. 

\section{Discussion and conclusion}\label{LAST}
In this paper we have given a fresh look on the determination of the $u$-plane integral (Coulomb branch integral) of the Donaldson-Witten theory on a four-manifold $X$ in the presence of a surface defect $\bfS$ inspired by \cite{Moore:1997pc, Tan:2009qq, Korpas:2017qdo}. 

We cosnidered the insertion of a specific $\overline{\CQ}$-exact operator $\tilde{\mathsf{I}}_{+}$ to the path integral of the low energy effective theory. This operator couples to the self-dual part of the curvature $\mathscr{F}$ of the extended bundle $\mathscr{E}\to X$. After some manipulations the ramified $u$-plane integral localizes to the cusps of the Coulomb branch.  As a result of our considerations, the determination of the ramified $u$-plane integral simplifies drastically since we do not need to use the techniques of lattice reduction anymore. The modularity of the integrand is preserved, as expected, and computation the ramified Donaldson invariants follows from a very simple formula. This comes in very close analogy to what was found in \cite{Korpas:2017qdo} where the usual Donaldson invariants were obtained (for specific K\"ahler surfaces) by a very similar simple formula as well and at the limit of vanishing volume for the embedded surface our result reduces to the one of the usual Donaldson-Witten theory. Reference \cite{Korpas:2017qdo} together with the present paper have shown that the relation between Donaldson-Witten theory and the theory of indefinite theta functions and mock modular forms is much stronger and deeper than what it was initially thought after the publication of the fundamental papers \cite{Gottsche:1996, Gottsche:1996aoa}. More generally, mock modular forms arise in increasing frequency in physical theories and appear to be of importance in low dimensional topology. See \cite{Cheng:2018vpl} for a recent relevant work on three-manifolds.

Returning to the $u$-plane integral, we would like to mention that it should not be so astonishing or surprising that a localization formula such as the one of Equation (\ref{contourInt}) appears in this context. This is a very generic feature of topological gauge theories. Similar integrals have often appeared in the literature for such theories.

An intresting direction to go forward would be to consider relating ramified $u$-plane integrals and indefinite theta functions for theories with higher rank gauge groups where the duality group lifts to ${\rm Sp}(2r,\BZ)$ where $r$ denotes the dimension of the Coulomb branch $\CB$. Some discussion towards this direction (for the usual Donaldson-Witten theory) has been presented in \cite{Korpas:2017qdo}.

Another direction worth of investigating is to generalize the results of this paper for four-manifolds that are not simply connected. Four-manifolds of the form $\BR \times Y$ (where $Y$ is a three-manifold), and $\BR^2 \times \Sigma_g$ (where $\Sigma_g$ is a genus $g$ Riemann surface) are of particular interest since they could relate our result, and especially mock modular forms, to the instanton Floer homology of $Y$ and the quantum cohomology of the moduli space of flat connection $\CM_{\rm flat}$ of $\Sigma_g$ respectively. See \cite{Gukov:2007ck} for a discussion of Donaldson-Witten theory and its reductions to such four-manifolds and the connections to surface operators.

As explained in the introduction, we do not expect that these tools will yield some new information about four-manifolds, at least not directly for ``conventional" operators and/or ``conventional" theories. The use of indefinite theta functions though, and mock modular forms more generally, in the world of topological gauge theories, might be useful towards finding new four-manifold invariants, e.g. by studying topological versions of superconformal theories \cite{Moore:2017cmm} and even topological class-$\CS$ theories. It is of great curiosity of ours to see if such tools can somehow be employed in the very much unexplored world of (maybe topological) $\CN=3$ theories and if such theories can provide any alternate roots for four-manifold invariants. Nevertheless, simplifying the evaluation of such Coulomb branch integrals, especially from the point of view of supersymmetric gauge theories, is quite important regardless of the mathematical problem of finding new invariants. 


\vspace{2em}

\noindent {\bf Acknowledgements}\\
We would like to thank Chris Aravanis, Dimitris Cardaris, Marcos Mari\~no, Jan Manschot, Gregory Moore, Sergey Mozgovoy, Stephen Pietromonaco, Samson Shatashvili and Meng-Chwan Tan for useful discussions, explanations and correspondence. 
\vspace{1em}


\appendix

\section{Surface operators, roots and characters} \label{App1}
In this appendix we would like to present some well known facts about surface operators and co-root lattices $\Lambda_{\rm cort.}$ that complements the discussion from Section \ref{Sec2}. 

Let us start with a remark. As we explained in Section \ref{Sec2} a surface operator is defined by prescribing a singular behavior for the gauge field along some surface $\bfS$. Nevertheless, there is another way to understand surface operators as a two dimensional theory supported on $\bfS$ whose flavor symmetry group is $G$, the gauge group of the four dimensional theory over $X$. Coupling the two dimensional theory to the four-dimensional one amounts to gauging $G$. For a concrete discussion see \cite{Gukov:2006jk, Gukov:2008sn, Gukov:2014gja}.

\begin{figure}[h]
\centering
\includegraphics[scale=1.4]{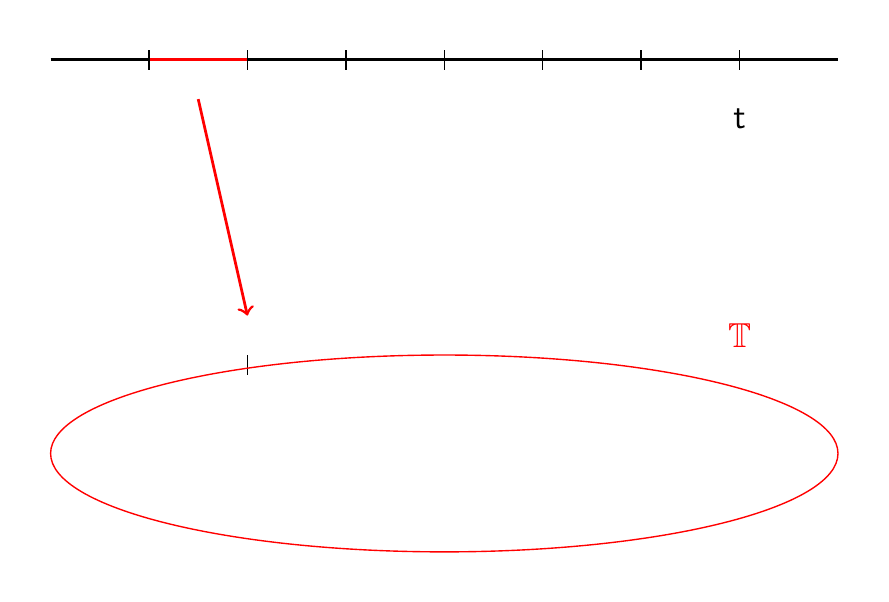}
\caption{The red segment in $\mathsf{t}$ corresponds to $\mathsf{t}/\pi_1(\mathbb{T})$. Each choice of segment corresponds to a different element of the fundamental group of $\mathbb{T}$. A lift from the base to $\mathsf{t}$ corresponds to the choice of a surface operator with $\alpha$ prescibed by the choice of line segment. \label{cochar}}
\end{figure}

In this paper we have focused on the approach of singularities of the gauge field $A$ and our task is to understand what we mean by lifting the bundle with connection one-form $A$ that is $\mathbb{T}$ valued to $\mathsf{t}$. Recall that to a semi-simple Lie group $G$ we associate a root lattice $\Lambda_{\rm rt.} \subset \mathsf{g}^{\vee}$. Similarly, for the Langlands dual group $^LG$ we associate a root lattice that is the co-root larrice of $G$, $\Lambda_{\rm cort.} \subset \mathsf{g}$. For simplicity, assume that $G$ is simply connected. Then the root lattice is embedded in the so-called character lattice $\Lambda_{\rm rt.} \subset  \Lambda_{\rm char.}$ which simply corresponds to ${\rm Hom}(\mathbb{T}, {\rm U}(1))$. Similarly, $\Lambda_{\rm cort.} \subset  \Lambda_{\rm cochar.}$ corresponds to ${\rm Hom}(\mathbb{T}^{\vee}, {\rm U}(1)) = {\rm Hom}( {\rm U}(1),\mathbb{T})$. The cocharacter lattice fits in the following exact sequence
\bes
0 \xrightarrow[\text{}]{\text{}} \Lambda_{\rm cochar.} \xrightarrow[\text{}]{\text{}} \mathsf{t} \xrightarrow[\text{}]{\text{}} \mathbb{T} \xrightarrow[\text{}]{\text{}} 0.
\ees
Actually it is possible to show that $\Lambda_{\rm cochar. }= \pi_1(\mathbb{T}) \cong \BZ^n$ where $n$ is the dimension of the Cartan subalgebra $\mathsf{t}$. This can be understood as follows. For any Lie group $G$ we can construct its universal cover $\tilde{G}$ and consider the following exact sequence
\bes
0 \xrightarrow[\text{}]{\text{}} \pi_1(G) \xrightarrow[\text{}]{\text{}} \tilde{G} \xrightarrow[\text{}]{\text{}} G \xrightarrow[\text{}]{\text{}} 0. 
\ees
In our case we can view the Cartan subalgebra $\mathsf{t}$ as the universal cover of the maximal torus $\mathbb{T}$ as follows
\be
0 \to \pi_1(\mathbb{T}) \xrightarrow[\text{}]{\text{$d$}} \mathsf{t} \xrightarrow[\text{}]{\text{exp}} \mathbb{T} \xrightarrow[\text{}]{\text{}} 0,
\ee
or equivalently we can view this configuration as principal $\pi_1(\mathbb{T})$-bundle over $\mathbb{T}$. A fiber of this bundle is exactly $\mathsf{t}$ as shown in the Figure \ref{cochar}. Therefore we have a natural identification of $\Lambda_{\rm cochar. }$ with $\pi_1(\mathbb{T})$.

Now let us present some connections of surface operators with Levi subgroups of $G$. The surface operators we have discussed are the simplest ones and belong to the so-called ``full" surface operators where $\alpha \in \mathbb{T} \cong {\rm U}(1)^n$. The classification of surface operators has been discussed in detail in \cite{Gukov:2006jk} (see also \cite{Cirafici:2013tna}) but let us repeat the main idea here. This classification consists of pairs $(\alpha, \BL)$ where $\alpha$ is the surface operators ``electric charge" as in the main part of this paper and $\BL$ is a Levi subgroup of $G$. Let us consider $G = {\rm SU}(k)$ for example. The Levi subgroups of $ {\rm SU}(k)$ are all possible groups of the form 
\bes
{\rm U}(1)^{l-1} \times \prod_{i=1}^l {\rm SU}(k_i).
\ees 
The minimal Levi subgroup of $ {\rm SU}(k)$ is its maximal torus $\mathbb{T}^{k-1}$ as in the main text of this paper, and the corresponding operator is a full surface operator. For $\BL =  {\rm SU}(k-1) \times  {\rm U}(1)$ the corresponding surface operator is called ``simple". Of course working with $ {\rm SU}(2)$ and $ {\rm SO}(3)$ restricts a lot the possibilities for surface operators that we can have in our theories. 

Finally let us mention that one can consider instead of $G$ its complexification $G_{\BC}$ and describe surface operators in terms of parabolic groups. This can be done for theories with $\CN \geq 2$ by combining the gauge field with the scalar field defining the complexified connection $\CA = A + i\phi$. The surface operators can be described by the the flat connection $\CA$ over a  $G_{\BC}$-bundle that along the embedded surface $\bfS$ its structure group $G_{\BC}$  is reduced to a parabolic subgroup $P \subset G_{\BC}$ \cite{Mehta1980}. This point is very useful for understanding surface operators in the context of six-dimensional Donaldson-Thomas theory \cite{Cirafici:2013tna}.
 
\section{Modular forms and theta functions}\label{App0} 
In this appendix we would like to collect some important notions from the theory of modular forms for the convenience of the reader. For a comprehensive exposition the reader is reffered to the plethora of available literature such as \cite{Serre, Zagier92, Bruinier08}.

\subsection{Modular groups}
The modular group $\operatorname{SL}(2,\mathbb{Z})$, is the group of integer matrices
with unit determinant
\be
\operatorname{SL}(2,\mathbb{Z})=\left\{ \left. \left( \begin{array}{ccc}
a & b   \\
c & d  \end{array} \right) \right| a,b,c,d\in \BZ; \, ad-bc=1\right\}.
\ee
which acts naturally on the Lobachevsky or upper half-plane $\BH = \{ \tau \in \BC ~|~ {\rm Im}(\tau) >0\}$ via 
\[
 \left( \begin{array}{ccc}
a & b   \\
c & d  \end{array} \right) \tau = \frac{a \tau+ b}{c\tau + d}.
\] 

We introduce moreover the congruence subgroup $\Gamma^0(n)$ 
\begin{equation}
\label{Gamma04} 
\Gamma^0(n) = \left\{  \left.\left( \begin{array}{ccc}
a & b   \\
c & d  \end{array} \right) \in \text{SL}(2,\BZ) \right| b = 0 \text{ mod } n \right\}.
\end{equation}
A holomorphic function $f: \BH \to \BC$ is a modular form of weight $k$ for any congruence subgroup $\Gamma \subset {\rm SL}(2,\BZ)$ if for any $\gamma \in \Gamma$ it satisfies 
\be
f(\gamma \tau) = (c\tau+d)^k f(\tau),
\ee
and it is holomorphic at the cusp at infinity $\tau \to i\infty$. In the following subsections we define various kinds of modular forms. There also exist mixed modular forms that are functions $f: \BH \times \bar{\BH}$ which transform as
\be 
f\left( \frac{a\tau+b}{c\tau+d}, \frac{a\sigma + b}{c\sigma+d} \right) = (c\tau+d)^k (c\sigma+d)^l f(\tau, \sigma).
\ee
The space of mixed modular forms for a modular subgroup ${\Gamma}$ is denoted as $\mathbb{M}_{(k,l)}(\Gamma)$.

\subsection{Eisenstein series}
We let $\tau\in \mathbb{H}$ and define $q=e^{2\pi i \tau}$. Then the Eisenstein series $E_k:\mathbb{H}\to \mathbb{C}$ for even $k\geq 2$ are defined as the $q$-series 
\be
\label{Ek}
E_{k}(\tau)=1-\frac{2k}{B_k}\sum_{n=1}^\infty \sigma_{k-1}(n)\,q^n,
\ee
with $\sigma_k(n)=\sum_{d|n} d^k$ the divisor sum. For $k\geq 4$, $E_{k}$ is a modular form of
$\operatorname{SL}(2,\mathbb{Z})$ of weight $k$. In other words, it transforms under $\operatorname{SL}(2,\mathbb{Z})$ as
\be
E_k\!\left( \frac{a\tau+b}{c\tau+d}\right)=(c\tau+d)^kE_k(\tau).
\ee
On the other hand $E_2$ is a quasi-modular form, which means that although it is a holomorphic function in the upper-half plane, the $\operatorname{SL}(2,\mathbb{Z})$ transformation of $E_2$ includes a shift in addition to the weight for any $\tau \in \BH$,
\be 
\label{E2trafo}
E_2\!\left(\frac{a\tau+b}{c\tau+d}\right) =(c\tau+d)^2E_2(\tau)-\frac{6i}{\pi}c(c\tau+d).
\ee
Eisenstein series $E_4(\tau)$ and $E_6(\tau)$ are somewhat special since they generate the ring of modular forms of ${\rm SL}(2,\BZ)$. On the other hand the ring of quasi-modular forms is generated by $E_2(\tau),E_4(\tau)$ and $E_6(\tau)$.

\subsection{Dedekind eta function}
The Dedekind eta function $\eta:\mathbb{H}\to\mathbb{C}$ is defined as
\be
\begin{split}
\eta(\tau) &=q^{\frac{1}{24}}\prod_{n=1}^\infty (1-q^n)\\
                 &=q^{\frac{1}{24}} (q)_{\infty}.
\end{split}
\ee
It is a modular form of weight $\frac{1}{2}$ under SL$(2,\BZ)$ with a
non-trivial multiplier system. It transforms under the generators of
SL$(2,\BZ)$ as
\be
\begin{split}
&\eta(-1/\tau)=\sqrt{-i\tau}\,\eta(\tau),\\
&\eta(\tau+1)=e^{\frac{\pi i}{12}}\, \eta(\tau). 
\end{split}
\ee

\subsection{Jacobi theta functions}
The classical Jacobi theta functions $\vartheta_j:\mathbb{H}\times
\mathbb{C}\to \mathbb{C}$, $j=1,\dots,4$, are defined as
\be
\label{Jacobitheta}
\begin{split}
&\vartheta_1(\tau,v)=i \sum_{r\in
  \mathbb{Z}+\frac12}(-1)^{r-\frac12}q^{r^2/2}e^{2\pi i
  rv}, \\
&\vartheta_2(\tau,v)= \sum_{r\in
  \mathbb{Z}+\frac12}q^{r^2/2}e^{2\pi i
  rv},\\
&\vartheta_3(\tau,v)= \sum_{n\in
  \mathbb{Z}}q^{n^2/2}e^{2\pi i
  n v},\\
&\vartheta_4(\tau,v)= \sum_{n\in 
  \mathbb{Z}} (-1)^nq^{n^2/2}e^{2\pi i
  n v}. 
\end{split}
\ee

We let $\vartheta_j(\tau,0)=\vartheta_j(\tau)$ for $j=2,3,4$.
Their transformations under the generators of $\Gamma^0(4)$ are 
\be
\label{Jacobitheta_trafos}
\begin{split}
&\vartheta_2(\tau+4)=-\vartheta_2(\tau),\qquad
\vartheta_2\!\left(\frac{\tau}{\tau+1}\right)=\sqrt{\tau+1}\,\vartheta_3(\tau),  \\
&\vartheta_3(\tau+4)=\vartheta_3(\tau),\qquad
\vartheta_3\!\left(\frac{\tau}{\tau+1}\right)=\sqrt{\tau+1}\,\vartheta_2(\tau),  \\
&\vartheta_4(\tau+4)=\vartheta_4(\tau),\qquad
\vartheta_4\!\left(\frac{\tau}{\tau+1}\right)=e^{-\frac{\pi
    i}{4}}\sqrt{\tau+1}\,\vartheta_4(\tau).  \\
\end{split}
\ee

\subsection{Siegel-Narain theta functions}
Siegel-Narain theta functions form a large class of theta functions for indefinite theta lattices. They only depend on the lattice data. The usual Jacobi theta functions are special case of the Siegel-Narain theta functions. We restrict to indefinite theta lattices of signarure $(1,n-1)$ and use the same definitions for the quadratic form $Q: H_2(X,\BZ) \to \BZ $ and the bilinear form $B: H_2(X,\BZ) \times H_2(X,\BZ)  \to  \BZ  $ as in (\ref{QUADBIN}). Furthermore we denote by $K$ the characteristic vector of the lattice $\Lambda$ such that for any vector $\bfv \in \Lambda$ we have $Q(\bfv) + B(\bfv , K) \in 2\BZ$. Then, given an element $J \in \Lambda \otimes \BR$ with positive norm, $Q(J)>0$, it is possible to decompose the space $\Lambda \otimes \BR$ to a positive definite subspace $\Lambda_{+} = \mathrm{span}\{ J \}$ as well as an orthogonal to it negative subspace $\Lambda_{-}$. The normalization of $J$ is defined as $\underline{J}:= \frac{J}{Q(J)}$ and we can use it to define projection of an arbitrary vector $\bfv$ to the positive and negative definite subspaces of $\Lambda$ as
\be
\begin{split}
\bfv_{+} :=& B(\bfv, \underline{J})\underline{J}, \\
\bfv_{-}:=& \bfv - \bfv_{+}. 
\end{split}
\ee
With the definitions given above, the Siegel-Narain theta function (series) that is of interest to the current paper and has appeared a few times in the main text is a map $\Psi_{\bfmu}^J : \BH \times \BC \to \BC$. The second argument of the map is called \emph{elliptic variable}. For a $J$ as the one discussed previously and for a conjugacy class $\bfmu \in \Lambda \otimes \BR$ the Siegel-Narain theta function reads
\be
\begin{split}
\Psi_{\bfmu}^{J}(\tau,\bfz) &= e^{-2\pi \tau_2 \bfb_{+}^2} \sum_{\bfv \in \Lambda + \bfmu} \partial_{\bar{\tau}} \Big(\sqrt{2\tau_2} B(\bfv + \bfb, \underline{J})  \Big)  \\
& \times (-1)^{B(\bfv, K)} q^{-\frac{Q(\bfv_{-})}{2}} \bar{q}^{\frac{Q(\bfv_{+})}{2}} e^{-2\pi i B(\bfv_{-},\bfz) - 2\pi i B(\bfv_{+}, \bar{\bfz})}
\end{split}
\ee
with $\bfb = \frac{\mathrm{Im}(\bfz)}{\tau_2} \in \Lambda \otimes \BR$. If we take $\bfb$ independent of $\bar{\tau}$ (as it usually is taken in the literature) then this theta function simplifies. In the current paper though this is not the case, actually $\bfz \in \BM_{(-1,0)}(\Gamma^0(4))$, and as a result $\bfb$ is not independent of the parameter $\bar{\tau} \in \bar{\BH}$ and furthermore we have $\partial_{\bar{\tau}}\bfb \in \BM_{(1,2)}(\Gamma^0(4))$. The modular properties of such theta functions are determined via the aid of Poisson resummation (as in the case of the standard Jacobi theta functions as well), for example see \cite{Borcherds:1996uda} for detailed formulas. In Section \ref{modularity} we gave some arguments on how the $T$ and $S$ transformations for the (ramified) Siegel-Narain series are derived but all details can be found in \cite[Appendix A]{Korpas:2017qdo}.

\section{Indefinite theta functions} \label{AppA}
\begin{figure}[h]
\includegraphics[scale=2.6]{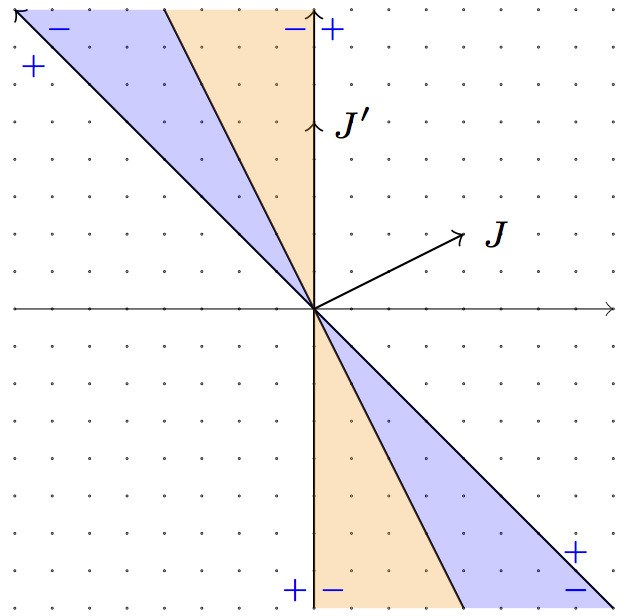}
\centering
\caption{The positive cones of some four-manifold $X$ defined for some (for illustrational purposes two-dimensional lattice) lattice $\Lambda$ of signature $(1,1)$. We have drawn vectors $J$ and $J'$ such that the latter has zero norm, $Q(J')=0$.  For these vectors only lattice points in the yellow are contribute to the sum of the indefinite theta functions.  \label{wallcrossing}}
\end{figure}

Indefinite theta functions (sometimes also called \emph{indefinite theta series}) are theta functions associated to an indefinite lattice $\Lambda$.  Such functions are special cases of mock modular forms, as mentioned in the introduction, and they have been getting a lot of attention since Zwegers' fundamental thesis \cite{Zagier92} (for a very recent exposition see \cite{Larry}). The relation of indefinite theta functions to the usual theta series (like the classical Jacobi theta functions) is very similar to the relation between mock modular forms and classical modular forms (see \cite[Section 3.3]{Manschot:2017xcr} for details). For our purposes we specialize to unimodular latices of signature $(1,n-1)$. It is clear that for such a lattice there will exist vectors that have negative definite norm and the sum, which can be divergent in general, schematically will read as
\[
\sum_{\bfv \in \Lambda} q^{-\pi \tau_2 \frac{\bfv^2}{2}}.
\]
Therefore we need to somehow regularize the sum such that we get a convergent series. This is done by summing only positive definite vectors with the ceveat that the series loses modularity properties. For the purposes of this paper, and for the quadratic form $Q$ and bilinear form $B$ as defined in (\ref{QUADBIN}), as well as $J,J' \in \Lambda \otimes \BR$ such that $B(J,J')>0$, $\underline{J}$ is the normalization of $J$, $\tau \in \BH$, $K$ a characteristic vector for $\Lambda$, $\bfz \in \Lambda \otimes \BC$, $\bfmu \in \Lambda \otimes \BR$ and $\bfb = \frac{1}{\tau_2}\mathrm{Im}(\bfz) \in \Lambda \otimes \BR$, the indefinite theta series is defined as 
 \be
 \begin{split} \label{BB1}
\Theta_{\bfmu}^{JJ'}(\tau,\bfz) := & \sum_{\bfv \in \Lambda + \bfmu} \frac{1}{2} \Big\{ \mathrm{sgn}(B(\bfv + \bfb, \underline{J})) - \mathrm{sgn}(B(\bfv + \bfb, \underline{J}'))  \Big\}  \\
& \times (-1)^{B(\bfv, K)} q^{-\frac{\bfv^2}{2}} e^{-2\pi i B(\bfv, \bfz)}.
\end{split}
\ee
This sum is convergent but does not transform as a modular modular as explained in \cite{ZwegersThesis}. Still, modular properties can be recovered by modifying slightly the kernel of the sum $\mathrm{sgn}(B(\bfv + \bfb, J)) - \mathrm{sgn}(B(\bfv + \bfb, J')) $. This modification amounts to adding to it some non-holomorphic terms. As it is explained in full detail in \cite{ZwegersThesis, MR2605321} there exists a modular completion $\widehat{\Theta}_{\bfmu}^{JJ'}$ of $\Theta_{\bfmu}^{JJ'}(\tau,\bfz)$. This amounts to substituting the sign functions of (\ref{BB1}) with rescaled error functions. The completion reads
\be
\begin{split} \label{BB2}
\widehat{\Theta}_{\bfmu}^{JJ'}(\tau,\bfz) := & \sum_{\bfv \in \Lambda + \bfmu} \frac{1}{2} \Big\{ E(B(\bfv + \bfb, \underline{J})) - E(B(\bfv + \bfb, \underline{J}'))  \Big\}  \\
& \times (-1)^{B(\bfv, K)} q^{-\frac{\bfv^2}{2}} e^{-2\pi i B(\bfv, \bfz)},
\end{split}
\ee
where, as explained in Section \ref{ThetaSection} as well, the (rescaled) error function is the map $E : \BR \to [-1,1]$ and it is defined as
\be
E(u) = 2 \int_0^u e^{-\pi t^2}dt = \mathrm{Erf}(\sqrt{\pi}u),
\ee
and note that when $\tau_2 \to \infty$ the function $E(u)$ from (\ref{BB2}) reduces to the sign function of (\ref{BB1}), that is
\be
\lim_{\tau_2 \to \infty} E(\sqrt{2\tau_2}u) = \mathrm{sgn}(u). 
\ee
Analytical continuation of $E$ (in order to be complex valued) makes it convergent only for 
\[
-\frac{\pi}{4} < \mathrm{Arg}(u) < \frac{\pi}{4}.
\]
In Figure \ref{wallcrossing} we give a graphical description of a lattice $\Lambda$ of signature $(1,1)$ and explain which points of $\Lambda$ have to be considered in sums such as (\ref{BB1}). 

The modular transformation properties of such indefinite theta functions under $\mathrm{SL}(2,\BZ)$ are explicitly derived in Zweger's thesis \cite[chapter 2]{ZwegersThesis} and also in \cite{Vigneras:1977} by  Vign\'eras. The generators $T$ and $S$ of $\mathrm{SL}(2,\BZ)$ act on $\widehat{\Theta}_{\bfmu}^{JJ'}(\tau,\bfz) $ as
\begin{equation}
\begin{split}
\widehat{\Theta}_{\bfmu + \frac{K}{2}}^{JJ'}(\tau+1,\bfz) &= e^{\pi i (\bfmu^2 - \frac{K^2}{4})}\widehat{\Theta}_{\bfmu + \frac{K}{2}}^{JJ'}(\tau,\bfz + \bfmu) \\[0.6em]
\widehat{\Theta}_{\bfmu + \frac{K}{2}}^{JJ'} \left( -\frac{1}{\tau}, \frac{\bfz}{\tau} \right) &= i(-i\tau)^{\frac{n}{2}} e^{-\frac{\pi i \bfz^2}{\tau} + \frac{\pi i K^2}{2}} \widehat{\Theta}_{ \frac{K}{2}}^{JJ'}(\tau, \bfz - \bfmu).
\end{split}
\end{equation}
As in \cite{Korpas:2017qdo} the object that we are quite interested in is the $\bar{\tau}$ derivative of the modular completed indefinite theta function \ref{BB2}. This derivative is exactly what we reffered to as the \emph{shadow} in the introductory section of this paper and its modular properties are much easier to determine than those of \ref{BB2} (although the notion of the \emph{shadow} is slightly different than the one used in \cite{ZwegersThesis} since the indefinite theta functions that appear here are mixed mock modular forms). In specific, we find that
\be
\partial_{\bar{\tau}} \widehat{\Theta}_{\bfmu}^{JJ'}(\tau, \bfz) = \Psi_{\bfmu}^J(\tau, \bfz) - \Psi_{\bfmu}^{J'}(\tau, \bfz)
\ee
where $\Psi_{\bfmu}^J$ is the Siegel-Narain function associated with $\Lambda$ and defined in Appendix \ref{App0}.  

If there exists a vector $\bfv_0 \in \Lambda$ such that $Q(\bfv_0) = 0$ then the modular completion of ${\Theta}_{\bfmu}^{JJ'}(\tau,\bfz)$ can be simplified because, for such type of lattices, we can choose vectors $J$ (and maybe also $J'$) such that they are identified with the vector $\bfv_0$. Then, as explained in \cite{ZwegersThesis}, the error function reduces to the sign function. Let us assume that there exists a vector $J' \in \Lambda$ such that $Q(J')=0$. The series will be convergent by further requiring that $B(\bfv + \bfb, J') \neq 0$ (we obviously cannot normalize $J'$ now) for any vector $\bfv \in \Lambda + \bfmu + \frac{K}{2}$ except if we also have that the other term in the kernel vanishes, i.e., if $B(\bfv + \bfb, J)=0$. The completion of  ${\Theta}_{\bfmu}^{JJ'}(\tau,\bfz)$ reads in that case  
\be
\begin{split} \label{RealTheta}
\widehat{{\Theta}}_{\bfmu}^{JJ'}(\tau,\bfz) =& \sum_{\bfv \in \Lambda + \bfmu + \frac{K}{2}} \frac{1}{2} \Big\{ E(B(\bfv + \bfb, \underline{J})) - \mathrm{sgn}(B(\bfv + \bfb, \underline{J}'))  \Big\}  \\
& \times (-1)^{B(\bfv, K)} q^{-\frac{Q(\bfv)}{2}} e^{-2\pi i B(\bfv, \bfz)},
\end{split}
\ee
the shadow of which exactly corresponds to a Siegel-Narain theta function, in specific we have
\be
\partial_{\bar{\tau}} \Theta_{\bfmu}^{JJ'}(\tau, \bfz) = \Psi_{\bfmu}^J(\tau, \bfz).
\ee
Finally, let us finish with a remark. It is important that $J'\in \Lambda$ since the modular complete function $\widehat{\Theta}_{\bfmu}^{JJ'}$ would not give a convergent series. An example of such a divergence is discussed in \cite[Appendix B.3]{Alexandrov:2017qhn}.


\bibliography{MyLibiCloud} 
\bibliographystyle{JHEP}

\end{document}